\documentclass[aps,prx,reprint,amsmath,amssymb,superscriptaddress,nofootinbib,floatfix,longbibliography]{revtex4-2}
\usepackage{graphicx,bm,mathtools}
\usepackage[colorlinks=true,linkcolor=blue,citecolor=blue,urlcolor=blue]{hyperref}
\usepackage{tikz}
\usepackage[subpreambles=true]{standalone}

\newcommand{\Tr}{\operatorname{Tr}}
\newcommand{\id}{\mathbb I}
\newcommand{\cL}{\mathcal L}
\newcommand{\D}[1]{\mathcal D_{#1}}
\newcommand{\ket}[1]{\lvert #1\rangle}
\newcommand{\bra}[1]{\langle #1\rvert}
\begin{document}

\newcommand{\tim}[1]{\textcolor{red}{\small {\textbf{(Tim:} #1\textbf{) }}}}

\newcommand{\new}[1]{\textcolor{blue}{\small {\textbf{(new:} #1\textbf{) }}}}

\newcommand{\yijian}[1]{\textcolor{cyan}{\small {\textbf{(yijian:} #1\textbf{) }}}}

\newcommand{\AN}[1]{\textcolor{orange}{\small {\textbf{(AN:} #1\textbf{) }}}}

\title{Instability of Markov chains under infinitesimal coherent dissipation}

\author{Amir-Reza Negari}
\affiliation{Perimeter Institute for Theoretical Physics, Waterloo, Ontario N2L 2Y5, Canada}
\affiliation{Institute for Quantum Computing, University of Waterloo, Waterloo, Ontario N2L 3G1, Canada}

\author{Yijian Zou}
\affiliation{Perimeter Institute for Theoretical Physics, Waterloo, Ontario N2L 2Y5, Canada}
\affiliation{Amazon Center for Quantum Computing, Pasadena, CA 91125, USA}\thanks{This work was completed prior to joining Amazon.}
\email{yijianzo@amazon.com}

\author{Timothy H. Hsieh}
\affiliation{Perimeter Institute for Theoretical Physics, Waterloo, Ontario N2L 2Y5, Canada}

\begin{abstract}
Under local Lindbladian evolution, the growth of conditional mutual information and its associated Markov length scale is not limited by Lieb-Robinson bounds, yet it has been unclear how quickly the Markov length can increase. We show that an arbitrarily short, strictly local Lindbladian can generate infinite Markov length from a product state. We also construct strictly local Lindbladians that converge rapidly to steady states with algebraically decaying conditional mutual information. Despite their long-range conditional correlations, the states we construct are local reversibly connected to a product state by sublinear depth circuits. Thus infinite Markov length does not imply slow relaxation nor obstruct phase stability when suitably defined.
\end{abstract}
\maketitle

\section{Introduction}
\label{sec:introduction}

Robustness to weak local perturbations is a defining feature of a phase of matter. For ground states of local Hamiltonians, the spectral gap expresses this robustness: a nonzero gap implies exponential clustering of correlations and, for broad classes of gapped systems, remains open under sufficiently weak local perturbations~\cite{HastingsKoma2006,BravyiHastingsMichalakis2010}. This links finite correlation length, stability under perturbations, and phase equivalence in the theory of quantum phases.

For mixed states there is often no underlying Hamiltonian, and therefore no obvious analogue of the gap. One recently proposed candidate~\cite{SangHsieh2025} is the \emph{Markov length}, the length scale governing the decay of conditional mutual information (CMI). For regions $A$ and $C$ separated by a buffer $B$, the CMI is defined as
\begin{equation}
 I(A:C|B)=S(AB)+S(BC)-S(B)-S(ABC).
 \label{eq:cmi}
\end{equation}
When this quantity decays exponentially with the width of $B$, the state has finite Markov length. Operationally, small CMI implies that information lost locally can be approximately recovered from a surrounding buffer. If the Markov length remains uniformly finite along a local dissipative evolution, the evolution can therefore be reversed by a quasi-local dissipative process~\cite{SangHsieh2025}. This relation between conditional correlations and recovery also underpins recent notions of mixed-state phase equivalence~\cite{SangEtAl2025LocalReversibility,Coser2019,mixedRG,Rakovszky2024} based on local reversibility and local stability.

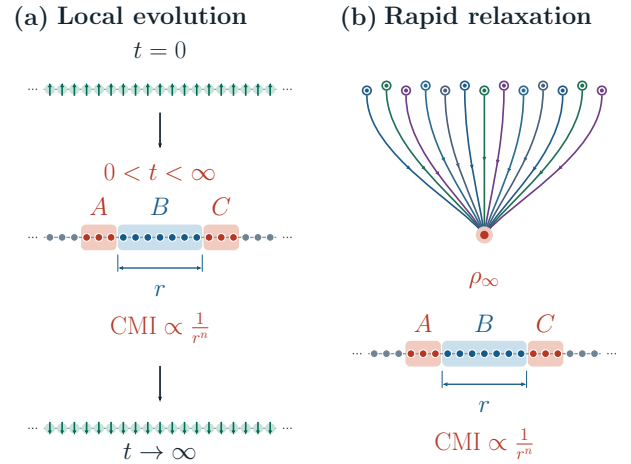
\begin{figure}[t]
\centering
\begin{tikzpicture}[x=1bp,y=1bp]
  \path[use as bounding box] (-6,-7) rectangle (528,408);
  \clip (-6,-7) rectangle (528,408);
  % Outer border: uncomment the next two lines to restore.
  % \draw[draw=black,line width=1bp,rounded corners=9bp]
  %   (-2,-3) rectangle (524,404);

  \node[heading] at (11,390) {(a)};
  \node[heading] at (47,390) {Local evolution};
  \node[heading] at (278,390) {(b)};
  \node[heading] at (314,390)
    {Rapid relaxation};
  % Panel divider: uncomment the next line to restore.
  % \draw[black,line width=.85bp] (263,17)--(263,387);

  % (a) Downward evolution between two product states.
  \node[time] at (131,348) {$t=0$};
  \SpinChain{131}{322}{1}
  \draw[flow] (131,303)--(131,273);
  \node[time,text=coral] at (131,248) {$0<t<\infty$};
  \StateChain{131}{201}
  \draw[flow] (131,96)--(131,66);
  \SpinChain{131}{45}{-1}
  \node[time] at (131,20) {$t\to\infty$};

  % (b) Thirteen initial states relax toward one stationary state.
  \foreach \i in {0,...,12}{%
    \pgfmathsetmacro{\xstart}{396+(\i-6)*16}
    \pgfmathsetmacro{\ystart}{322+(mod(\i,2)==1 ? 3 : -1)}
    \pgfmathsetmacro{\xcontrol}{396+.18*(\xstart-396)}
    \pgfmathtruncatemacro{\colorindex}{mod(\i,5)}
    \pgfmathsetmacro{\markposition}{.48+mod(\i,3)*.035}
    \draw[draw=traj\colorindex,line width=1.35bp,
      decoration={markings,mark=at position \markposition with
        {\arrow{Latex[length=4.5bp,width=3.4bp]}}},postaction={decorate}]
      (\xstart,{\ystart-4})
      .. controls (\xstart,{\ystart-52}) and (\xcontrol,227) .. (396,203);
    \filldraw[fill=white,draw=traj\colorindex,line width=1.35bp]
      (\xstart,\ystart) circle[radius=3.5bp];
    \fill[traj\colorindex] (\xstart,\ystart) circle[radius=1.65bp];
  }
  \fill[corallight] (396,203) circle[radius=7bp];
  \fill[coral] (396,203) circle[radius=3.4bp];
  \node[text=coral,font=\FigureMathFont] at (396,164) {$\text{}\rho_\infty$};
  \StateChain{396}{106}
\end{tikzpicture}
\caption{
\textbf{Infinite Markov length under local dissipation.}
(a) We construct a local Lindbladian acting on a product state of spins which
produces algebraically decaying CMI for all intermediate states
at any finite $t>0$. 
(b) We construct a local Lindbladian which rapidly relaxes to a stationary
state $\rho_\infty$ with algebraically decaying CMI.
}
\label{fig:summary}
\end{figure}
The analogy with the spectral gap raises the question: \emph{is finite Markov length itself stable under weak local dynamics?} This question differs from Ref.~\cite{SangHsieh2025}, which shows what follows (i.e. phase equivalence) once finite Markov length along an evolution is assumed. 
% The recovery construction of Ref.~\cite{SangHsieh2025} requires finite Markov length both along the evolution and in partially evolved intermediate states; the expectation that a sufficiently small local change cannot suddenly turn a finite Markov length into an infinite one remains an assumption. 
For gapped local Hamiltonians, there are fine-tuned cases in which the gap can close under an infinitesimal extensive local perturbation to the Hamiltonian\footnote{We thank Dominic Else and Shengqi Sang for discussions of this.}.  Can Markov length also diverge under an infinitesimal perturbation? Conditional correlations are not subject to Lieb-Robinson light cones and can become long-ranged under sufficiently strong ($O(1)$ time) decoherence \cite{PhysRevLett.133.200402, ZhangGopalakrishnan2024}.  For certain incoherent channels acting on Gibbs states of stabilizer Hamiltonians, stability of finite Markov length has been established~\cite{ZhangGopalakrishnan2025Stability}. Nonetheless, it has been unclear whether finite Markov length is generally robust or can be unstable to arbitrarily weak local dissipative perturbation.

We show that the latter can occur. Starting from a product state, a strictly local Lindbladian evolution can cause the Markov length to immediately diverge. We analyze this dynamics exactly in a fermionic model and use it to construct a strictly local spin model with the same phenomena. At any fixed $t>0$, the CMI has an algebraic lower bound as the buffer grows, ruling out any finite Markov length. 
% At long times, however, the evolution approaches another product state. 
An infinitesimal amount of local dissipation can therefore take a state from zero to infinite Markov length.  We further show the same instability appears in nonlinear correlations and in the canonical purification of the state.
%along a path with two completely uncorrelated endpoints.

Using these examples, we then construct a class of strictly local Lindbladians which rapidly relax to stationary states with infinite Markov length.  In this class, an arbitrarily weak local perturbation of a generator preparing a product state can produce a stationary state with infinite Markov length while relaxation remains rapid.  This complements recent results showing that rapid mixing together with frustration freeness of a local Lindbladian imply exponential decay of CMI~\cite{RosaRuizEtAl2026}: our examples are rapidly mixing but not frustration free, so rapid relaxation alone does not protect finite Markov length. 

Regarding the definition of mixed-state phases~\cite{SangEtAl2025LocalReversibility,Coser2019,mixedRG, Rakovszky2024}, our examples further highlight subtleties which were discussed in \cite{ChenHierarchical2026,recent}.  We show that, although algebraic CMI obstructs a locally reversible connection to a product state with error superpolynomially small in system size $L$, there exists such a locally reversible connection with sublinear in $L$ locality and $\mathrm{poly}(L^{-1})$ error.  Diverging Markov length alone should therefore not be taken as evidence for slow dynamics or an obstruction to dissipative connection with a trivial state.  
% This instability limits what a divergent Markov length implies about phase connectivity. 
% Algebraic CMI obstructs a polylogarithmically local recovery against local erasure reversible connection exponentially accurate recovery of a locally erased region from its surrounding buffer. In this sense, the local recovery structure associated with finite Markov length is lost. The same states nevertheless remain connected to a product state in both directions by local dissipative dynamics. For the trajectory we construct, an exponentially local evolution also retraces the state family toward the initial product state. Two-way dissipative connectivity is therefore weaker than the local reversibility encoded by finite Markov length. This distinction bears on recent definitions of mixed-state phases based on locally reversible channels~\cite{SangEtAl2025LocalReversibility,LiFirankoHsieh2026}: failure of finite Markov length need not, by itself, obstruct more general dissipative conversion.

% These results qualify the gap analogy. Finite Markov length remains a \emph{sufficient condition} for local recovery and reversibility, but it is not stable under arbitrarily weak local dissipation. 

We first study these phenomena in free fermion systems, where we can control the dynamics and information-theoretic quantities analytically, then transform them into spin systems through a construction that preserves locality and the relevant CMI. We use this construction first for a dissipative trajectory with an immediate divergence of Markov length and then for rapidly relaxing local Lindbladians whose stationary states have infinite Markov length. We then examine local recovery, dissipative reversibility, and related nonlinear correlations, and discuss the implications for mixed-state quantum phases.

\section{Immediate divergence of Markov length}
\label{sec:divergence}

Here we first construct an exactly solvable fermionic evolution that develops infinite Markov length immediately from a product state. The construction exploits the nonanalytic dependence of entropy for Gaussian states on occupation eigenvalues near zero, and this nonanalyticity produces subleading corrections to the entropy that determine the CMI. 
% We choose a local dissipation whose occupation spectrum reaches zero at every finite positive time, so that this nonanalyticity produces algebraic corrections to the entropy that survive in the CMI. 
% We explain this mechanism in Sec.~\ref{sec:fermions}. 
Its free-fermion structure lets us evaluate the CMI efficiently at large distances and prove an algebraic decay lower bound at every finite time. We then map the evolution to a strictly local spin Lindbladian, preserving the CMI of consecutive intervals.

\subsection{Fermionic construction}
\label{sec:fermions}
\label{sec:finite-time}

We consider a chain of complex fermions initially in the vacuum $\ket{\Omega}$, evolving under the following dissipation: 

\begin{equation}
\dot\rho=\sum_j\D{F_j}(\rho),\qquad
F_j=c_j^\dagger+c_{j+1}^\dagger .
\label{eq:model}
\end{equation}
Here $\D{F}(\rho)=F\rho F^\dagger-\{F^\dagger F,\rho\}/2$, and the rate and lattice spacing set the units of time and distance. For consecutive intervals $A,B,C$ of lengths $a,b,c$, we study the CMI as the separating buffer $B$ grows.

The source of the long-range CMI is visible at short times. Writing $\ket j=c_j^\dagger\ket\Omega$, each jump creates the state $F_j\ket\Omega=\ket j+\ket{j+1}$. To first order in time, Supplemental Secs.~\ref{app:fermion-response} and \ref{app:response-asymptotics} shows the nontrivial block of the reduced density matrix on a contiguous interval is $tK$, where $K_{ij}=2\delta_{ij}+\delta_{i,j+1}+\delta_{i,j-1}$. The infinite-chain momentum basis eigenvalues are $\kappa(k)=|1+e^{ik}|^2=2+2\cos k$, which vanishes quadratically at $k=\pi$. This vanishing gap enters the entropy through the nonanalytic function $-\kappa\log\kappa$. After the extensive contribution cancels in the CMI, the remaining finite-size correction gives $\left.\partial_t I(A:C|B;t)\right|_{t=0^+}=3\zeta(3)ac/b^4+O(b^{-5})$ for fixed $a,c$. Supplemental Secs.~\ref{app:fermion-response} and \ref{app:response-asymptotics} give the detailed derivation.

This short-time response suggests an immediate loss of finite Markov length. The free-fermion structure lets us study the behavior at fixed $t>0$ and arbitrarily large $b$. The jumps are linear in fermion operators, so the state remains Gaussian. Since $\langle c_i c_j\rangle=0$, the covariance $N_{ij}(t)=\langle c_i^\dagger c_j\rangle$ completely determines the evolving state~\cite{Peschel2003,BarthelZhang2022}. Due to translation invariance, one can transform to momentum space $n(k,t) := L^{-1}\sum_{m,n} e^{ik(m-n) }N_{mn}(t)$. It obeys $\dot n(k,t)=\kappa(k)[1-n(k,t)]$, giving the exact solution $n(k,t)=1-e^{-t\kappa(k)}$ and $N(t)=\id-e^{-tK}$.

\begin{figure}[tbp]
\centering
\includegraphics[width=0.96\columnwidth]{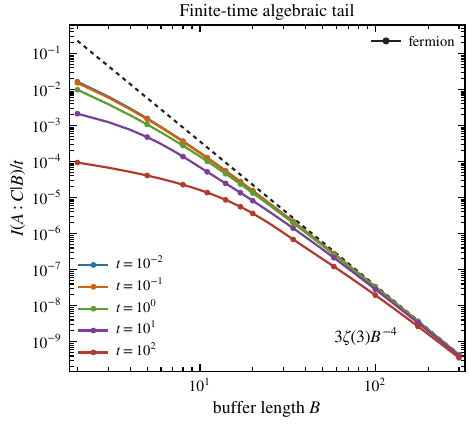}
\caption{CMI with single-site end regions for the fermionic Gaussian evolution at various times.
% divided by time for the fermionic Gaussian evolution at
% $t=10^{-2},10^{-1},1,10,10^2$, with single-site end regions.
% The buffer length is denoted $B$ in the figure and $b$ in the text.
The dashed line is $3\zeta(3)B^{-4}$. At large buffers, the data are
consistent with the asymptotic form in Eq.~\eqref{eq:predicted-cmi}.
% The bosonic comparison is shown in Supplemental
% Fig.~\ref{fig:gaussian-comparison}.
}
\label{fig:gaussian}
\end{figure}

The exact covariance also allows efficient numerical evaluation of the finite-time CMI. For an interval of $L$ sites, we restrict the infinite-chain covariance to obtain $N_L(t)$. If $n_\alpha$ are its eigenvalues, then $S_L(t)=\sum_{\alpha=1}^L h(n_\alpha)$, with $h(n)=-n\log n-(1-n)\log(1-n)$. Combining the entropies for intervals of lengths $a+b$, $b+c$, $b$, and $a+b+c$ gives the CMI. We can therefore reach buffer sizes far beyond those accessible to direct many-body calculations.

Figure~\ref{fig:gaussian} shows that the algebraic behavior persists at every finite time. For single-site end regions, the curves of $I/t$ at different times approach the same $b^{-4}$ dependence at large buffer size, consistent with
\begin{equation}
I(A:C|B;t)\sim\frac{3\zeta(3)tac}{b^4},
\label{eq:predicted-cmi}
\end{equation}
for fixed $a,c$. 
% \yijian{for small $t$?} \AN{this works for any $t$, since here we are first taking $b$ large limit}

The exact covariance also lets us prove analytically that the decay cannot become exponential at any finite positive time. Supplemental Sec.~\ref{app:finite-time} shows that, for single-site end regions ($a=c=1$), $I(A:C|B;t)\geq C(t)/(b+1)^4$ for sufficiently large $b$, with $C(t)>0$ for every finite $t>0$. %Time is held fixed while the buffer is taken large. 
This bound rules out exponential CMI decay with any finite length scale. The Markov length therefore becomes infinite immediately and remains infinite at any $t>0$. The argument also applies when $A$ is an interval and $B$ is two intervals surrounding $A$. 

This is compatible with the long-time limit. As $t\to\infty$, every finite region approaches the filled product state and its CMI tends to zero. The evolution therefore connects two product states, while every state at finite nonzero time between them has infinite Markov length.
% A bosonic model with jumps $a_j^\dagger+a_{j+1}^\dagger$ and vacuum input gives a closely related example. It has the same short-time CMI response and remains Gaussian. The numerical results in Supplemental Fig.~\ref{fig:gaussian-comparison} are consistent with the same fixed-time algebraic tail. 
We now use the fermionic construction to obtain a strictly local spin evolution with the same CMI behavior.

\subsection{Spin construction}
\label{sec:spins}

We now transfer the fermionic result to a strictly local spin system. A direct Jordan--Wigner transformation ($c_j^\dagger=\left(\prod_{\ell<j}Z_\ell\right)\sigma_j^-$
where $\sigma^-=(X-iY)/2$) is insufficient: a single fermion creation operator maps to a spin operator with a parity string extending to the boundary. We remove this string by introducing one maximally mixed auxiliary fermion per site and replacing each jump $F_j$ by $\widetilde F_j=i\gamma_jF_j$, where $\gamma_j$ is a Majorana operator of the auxiliary mode. The product $\widetilde F_j$ is even in fermion parity, so alternating the physical and auxiliary modes cancels the Jordan--Wigner strings outside a finite region.

The auxiliary modes leave the physical evolution unchanged. For physical states even in fermion parity, the dissipator generated by $\widetilde F_j$ reproduces that of $F_j$, while the auxiliary modes remain maximally mixed at all times. To see this, let $\widehat\rho$ denote the physical state embedded with
maximally mixed auxiliary fermions. It is an even polynomial
of physical Majoranas and commutes with $\gamma_j$, giving
$\widetilde F_j\widehat\rho\,\widetilde F_j^\dagger
=F_j\widehat\rho F_j^\dagger$
and
$\widetilde F_j^\dagger\widetilde F_j=F_j^\dagger F_j$.
Thus the lifted dissipator preserves this embedding at every
time. After the Jordan--Wigner transformation, the two spin
sublattices can nevertheless be correlated, even though the
auxiliary marginal remains maximally mixed. Supplemental Sec.~\ref{app:spins} gives the operator identities and explicit fermion-to-spin mapping. Related auxiliary-Majorana constructions have been used to obtain local spin representations of fermionic systems~\cite{VerstraeteCirac2005}; here the maximally mixed auxiliaries preserve the system's full dissipative trajectory.

With the physical qubits at sites $2j$ and the auxiliaries at $2j+1$, the spin state evolves under $\dot\rho_{\rm spin}=\sum_j\D{J_j}(\rho_{\rm spin})$, with
\begin{equation}
J_j=\sigma^-_{2j}X_{2j+1}
+iY_{2j+1}\sigma^-_{2j+2},
\label{eq:spin-jump}
\end{equation}
where $\sigma^-=(X-iY)/2=\ket\downarrow\bra\uparrow$. Each jump acts on only three consecutive qubits. The initial state is mapped to a product state $\prod_j\left(  |\uparrow\rangle\langle\uparrow|_{2j} \otimes\frac{\mathbb{I}_{2j+1}}{2}\right)$. 
%The fermionic vacuum maps to a product state with the physical spins polarized up and the auxiliary spins maximally mixed, so the auxiliaries introduce no initial correlations.

To compare the CMI, we group each physical qubit and its auxiliary into one cell. For any consecutive interval $R$ of complete cells, the auxiliaries contribute only an extensive entropy, $S_{\rm spin}(R;t)=S_{\rm fermion}(R;t)+|R|\log2$. This contribution cancels from the CMI, giving
\begin{equation}
I_{\rm spin}(A:C|B;t)
=
I_{\rm fermion}(A:C|B;t)
\label{eq:spin-cmi}
\end{equation}
for consecutive regions $A,B,C$. The numerical results of Fig.~\ref{fig:gaussian} and the finite-time lower bound of Sec.~\ref{sec:fermions} therefore carry over to the spin system, with distances measured in cells. Supplemental Sec.~\ref{app:spins} discusses the corresponding shielding geometry on an open chain.  At long times, every finite region approaches another product state: the physical spins become polarized down while the auxiliaries remain maximally mixed. 
% The evolution therefore connects two product states related by onsite spin flips, yet every intermediate state at finite $t>0$ has infinite Markov length. 
% We next ask whether infinite Markov length can also occur in stationary states of rapidly relaxing local dynamics.

\section{Steady states with rapid relaxation}
\label{sec:preparation}
\label{sec:rapid}

Infinite Markov length can thus appear immediately along a dissipative trajectory. We now show that such long-range CMI can arise in the steady state of a rapidly relaxing local Lindbladian. Our strategy is to build the mechanism of Sec.~\ref{sec:fermions} into the stationary state. We design a local Lindbladian whose stationary occupation spectrum approaches zero, so that the same entropy singularity produces algebraic CMI, while the dynamics remains rapidly relaxing. 
% We construct strictly local Lindbladians that relax rapidly to stationary states with infinite Markov length.

\subsection{Fermionic steady states}

We construct this family of Lindbladians by combining two local cooling processes. Define the Bogoliubov modes $d_j=(c_j+c_j^\dagger+c_{j+1}-c_{j+1}^\dagger)/2$, which obey the canonical fermionic anticommutation relations. For $0<p<1$, we take
\begin{equation}
\mathcal L_p=(1-p)\sum_j\D{c_j}+p\sum_j\D{d_j}.
\label{eq:bogoliubov-lindbladian}
\end{equation}
The first term cools the physical fermions
toward the trivial product vacuum; the second cools
Bogoliubov quasiparticles on neighboring sites
toward the topological superconducting ground state of
the Kitaev chain at its fully dimerized point.
Each process favors a different vacuum, and their competition
produces a mixed stationary state. Related dissipative
preparation schemes using Bogoliubov cooling were developed
in Refs.~\cite{DiehlEtAl2011,BardynEtAl2013}.

The dynamics can be solved exactly. Using a Grassmann generating function for Majorana correlation functions, we show in Supplemental Sec.~\ref{app:bogoliubov-parent} that $\mathcal L_p$ has a unique Gaussian stationary state $\rho_\infty$ and that all dependence on the initial state decays exponentially. The relaxation rate remains finite uniformly in $p$ and system size. Under this evolution, the density matrix on $M$ sites approaches the corresponding marginal of $\rho_\infty$ with error $\varepsilon$ in trace norm in $O[\log(M/\varepsilon)]$ time. More explicitly, for an interval $R$ of $M$ sites, $\|\rho_R(t)-\rho_{\infty,R}\|_1\leq\pi M e^{-t/2}$. The intuition is that both cooling processes damp covariance deviations at the same rate, giving $\partial_t\Gamma=\Gamma_p-\Gamma$, where $\Gamma_p$ is the stationary covariance matrix. 

The stationary covariance has a simple form. With $\eta_j=c_j+c_j^\dagger$ and $\xi_j=-i(c_j-c_j^\dagger)$, its only nonzero elements, apart from antisymmetry, are
\(
(\Gamma_p)_{\eta_j,\xi_\ell}
=-(1-p)\delta_{j\ell}-p\delta_{\ell,j+1}.
\label{eq:bogoliubov-real-covariance}
\) The stationary covariance is a weighted average of those
of the trivial product vacuum and the topological superconducting
ground state of the Kitaev chain.
Despite this short-ranged covariance, the CMI has algebraic decay. Writing $q_p(k)=1-p+pe^{ik}$, the smaller occupation
eigenvalue is $x_p(k)=(1-|q_p(k)|)/2=p(1-p)k^2/4+O(k^4)$.
Indeed, $1-|q_p(k)|^2=p(1-p)(2-2\cos k)$ is proportional
to the discrete-Laplacian symbol. The entropy therefore
contains the same $k^2\log|k|$ nonanalyticity as in
Sec.~\ref{sec:fermions}. Supplemental Sec.~\ref{app:bogoliubov-cmi} gives, for single-site $a=c=1$ end regions and fixed $0<p<1$,
\begin{equation}
I(A:C|B)_{\rho_\infty}
\sim
\frac{3\zeta(3)p(1-p)}{4b^4}.
\label{eq:bogoliubov-cmi-asymptotic}
\end{equation}
Thus $\rho_\infty$ has infinite Markov length throughout the family, despite being the unique stationary state of rapidly relaxing local dynamics. 

\subsection{Spin steady states}

The auxiliary construction of Sec.~\ref{sec:spins} also yields a rapidly converging, strictly local spin model with power-law CMI in the steady state. Let $\gamma_j,\delta_j$ denote the Majorana operators of an auxiliary fermion at site $j$. We introduce cooling jumps $i\gamma_j c_j$ and $i\gamma_j d_j$ with rates $1-p$ and $p$, respectively, together with the jumps $-i\gamma_j\delta_j$ and $i\gamma_j\gamma_{j+1}$, both at unit rate. The auxiliary noise $-i\gamma_j\delta_j$ and $i\gamma_j\gamma_{j+1}$ dephases the occupation degrees of freedom and drives the auxiliary modes to the maximally mixed state. Consequently, the effective evolution of the system together with the auxiliary fermions reproduces the same Gaussian evolution on the system.

Because all jump operators involve an even number of local fermionic operators, the Jordan--Wigner transformation maps the dynamics to strictly local spin dissipators. The corresponding spin jumps are $\sigma^+_{2j}X_{2j+1}$ and $(-Y_{2j}X_{2j+1}+iY_{2j+1}Y_{2j+2})/2$, together with the auxiliary jumps $Z_{2j+1}$ and $Y_{2j+1}Z_{2j+2}X_{2j+3}$, where $\sigma^+=(X+iY)/2$. In Supplemental Sec.~\ref{app:spin-steady-parent}, we prove in detail that, on any local interval of size $L$, this spin dynamics has a unique stationary state and converges to it within error $\varepsilon$ in time $O[\log(L/\varepsilon)]$. Moreover, its CMI is equal to that of fermionic $\rho_\infty$ and therefore exhibits the same algebraic decay.

One important distinction from the purely fermionic construction of the previous section is that the spin dynamics inherits a parity symmetry. In the fermionic representation, the total fermion parity commutes with all jump operators, so a global charge remains conserved throughout the evolution. This conserved quantity is invisible to local degrees of freedom, and the different global stationary states are locally indistinguishable.

Reference~\cite{RosaRuizEtAl2026} proves exponential CMI decay
for the unique stationary states of rapidly mixing,
frustration-free Lindbladians with finite-range spin interactions.
For CMI on a finite region $ABC$, with $B$ surrounding $A$,
their proof can be adapted to require only uniform rapid
convergence of local reduced states to common stationary
marginals from every initial state, so global uniqueness
is unnecessary. Our construction shows why frustration freeness,
however, remains crucial. A frustration-free Lindbladian leaves
its stationary state unchanged under each local contribution
separately. Equation~\eqref{eq:bogoliubov-lindbladian} does not
have this property, since $\rho_\infty$ is stationary only
under the combined action of the local cooling terms.
Therefore, algebraic CMI here rules out frustration-free local
parents with uniform rapid local relaxation for these states.

The recovery picture helps explain the role of frustration freeness. After erasing a central region, rapid mixing restores the state under the full evolution. To obtain a local recovery map, we restrict this evolution to the erased region and its surrounding buffer. For a frustration-free evolution, each local term leaves the stationary state invariant, so adding or removing such terms does not disturb it. In particular, removing terms at the buffer boundary leaves the original state stationary. Together with rapid mixing, locality then allows the restricted evolution to approximate the full recovery with an error exponentially small in the buffer width, giving finite Markov length. In our model, however, removing boundary-crossing terms changes the state at the cut, which obstructs the local recovery argument.

This family also shows the instability of finite Markov length to weak perturbations of a trivial Lindbladian. At $p=0$, Eq.~\eqref{eq:bogoliubov-lindbladian} reduces to onsite loss, whose stationary state is the product vacuum. Turning on an arbitrarily small $p$ changes the generator only perturbatively while relaxation remains uniformly fast; nevertheless, the stationary Markov length becomes infinite for every $p>0$. Evolution under $\mathcal L_p$ also rapidly takes the stationary state at $p'$ toward that at $p$ locally, with a rate independent of the total system size (Supplemental Sec.~\ref{app:bogoliubov-parent}). Under the dynamical notion of stability based on rapid relaxation between stationary states of nearby generators~\cite{Rakovszky2024}, the above family of states are all in the same phase.

\section{Implications for mixed-state phases}
\label{sec:implications}

Mixed-state phase equivalence can be defined through two-way connectivity by local channels \cite{Coser2019}. A more restrictive definition requires the connecting evolution to be locally reversible gate by gate~\cite{SangEtAl2025LocalReversibility}. This condition ensures the preservation of topological degeneracy and symmetry anomalies~\cite{SangEtAl2025LocalReversibility}, as well as superpolynomial decay of CMI~\cite{Yi2026}.

If one allows for only superpolynomially small error, then our example shows that the local reversibility notion of phase equivalence is not stable under generic weak local dissipation. In Supplemental Sec.~\ref{app:approx-local-reversibility}, we extend the CMI invariance result of Ref.~\cite{Yi2026} to circuits with a sublinear total light cone and superpolynomially small total reversal and preparation errors. These circuits preserve superpolynomially small CMI across buffers proportional to the system size. Our evolution, however, generates an algebraic CMI lower bound at every fixed $t>0$, starting from a product state. It therefore leaves the product state's locally reversible equivalence class immediately, even when sublinear light cones are allowed. 
% This class consequently does not contain a neighborhood of the product state that is stable under these weak local dissipative perturbations.

Two aspects of the definition matter for this conclusion. The first is the requirement of gate-by-gate reversal. Our trajectory can still be retraced by a local evolution with exponentially decaying couplings, following $\rho_{\rm rev}(s)=\rho_{t_*e^{-s}}$ and approaching the initial product state as $s\to\infty$ (Supplemental Sec.~\ref{app:return}). This gives a local reversal of the full trajectory, without undoing the forward gates individually.

The second is the required accuracy. If inverse-polynomial errors are allowed in both local reversal and phase equivalence, states distinguished when demanding superpolynomial accuracy can now become connected, as illustrated by the hierarchical construction of Ref.~\cite{ChenHierarchical2026} and certain pure states\footnote{For pure states,
uniformly gapped paths admit local-circuit approximations compatible with superpolynomial accuracy.  On the other hand, polynomial accuracy allows states with 
power-law correlations to be connected to product states; Appendix~\ref{app:superpolynomial-accuracy}
gives a $W$-state type example. This example admits a depth-zero
preparation with polynomial error, while superpolynomial accuracy
requires at least linear circuit complexity.}. The recovery construction of Ref.~\cite{SangHsieh2025} also shows how sufficiently fast power-law CMI decay can allow gate-by-gate reversal. For each forward gate on $A$, one constructs a Petz recovery map on $A$ and a surrounding buffer $B$. The map restores the state before the gate up to an error controlled by $\sqrt{I(A:C|B)}$, where $C$ is the rest of the system. Applying these maps in reverse order then reverses the circuit, with the errors adding over the gates. If $I(A:C|B)=O(b^{-\alpha})$ holds uniformly over the partially evolved states, each recovery error is $O(b^{-\alpha/2})$.  Choosing $b=L^\beta$, the total error for $O(L)$ gates is $O(L^{1-\beta\alpha/2})$.  Hence, if $\alpha>2$, there is a $\beta<1$ giving a two-way locally reversible connection with inverse-polynomial total error $O(L^{1-\beta\alpha/2})$ and a sublinear light cone.

\section{Nonlinear correlations}
\label{sec:nonlinear}

A similar instability appears in nonlinear probes of mixed states.
Correlators involving fractional powers of $\rho$ are used as
diagnostics of mixed-state phases, including fidelity and Wightman
correlators for strong-to-weak spontaneous symmetry breaking and
related phenomena~\cite{LessaEtAl2025,LiuEtAl2025,LiFirankoHsieh2026}.
In our model in Sec.~\ref{sec:fermions},
ordinary correlation functions remain short ranged due to
Lieb-Robinson bounds, while the nonlinear correlator
$W_t(r)=\Tr(\sqrt{\rho_t}\,n_0\sqrt{\rho_t}\,n_r)
-\langle n_0\rangle\langle n_r\rangle$
behaves as $W_t(r)=t/(\pi^2r^4)+O_t(r^{-6})$ for every $t>0$.
Equivalently, the canonical purification develops an algebraic
correlation immediately. Supplemental Secs.~\ref{app:nonlinear}
and \ref{app:fidelity-correlations} give the calculation and its
connection to fidelity correlators.

% \tim{would be good to also mention the other example where canonical purification is short-range correlated} \AN{I fixed this too} \tim{thanks. also maybe somewhere around here, you can comment on Renyi CMIs being exponentially decaying, even for the original (above) example, correct?} \AN{I added a few sentences at the end of the paragraph explaining this, using the analyticity argument again.}

Algebraically decaying CMI can also occur while the canonical
purification remains short ranged. For example, if we consider
instead the jump operators
\(F_j=\frac{\sqrt3}{8}
(c_j^\dagger-2c_{j+1}^\dagger+c_{j+2}^\dagger)\) and
\(G_j=\frac18(3c_j+6c_{j+1}-c_{j+2})\), then
$n(k,t)=q_t(2-2\cos k)^2$, where $q_t=3(1-e^{-t})/64$.
Near $k=0$, the occupation vanishes as $n(k,t)\sim q_t k^4$,
so the purification kernel
$\sqrt{n(k,t)[1-n(k,t)]}\sim\sqrt{q_t}\,k^2$
is analytic. Its Fourier transform then decays exponentially,
so both the physical state and its canonical purification have
short-range correlations. The entropy contribution
$-n(k,t)\log n(k,t)$, however, retains a $k^4\log|k|$
nonanalyticity. Supplemental Sec.~\ref{app:short-range-purification}
proves a power-law $b^{-6}$ CMI lower bound
for every $t>0$, showing that the Markov length diverges
immediately even though the canonical purification remains
short ranged.

We also expect algebraic decay in nonlinear correlators involving
$\rho^{1/4}$ (such as the connected correlator
$\Tr(\rho_t^{1/4}n_0\rho_t^{3/4}n_r)
-\langle n_0\rangle\langle n_r\rangle$),
since the fourth-root factor
$n(k,t)^{1/4}\sim q_t^{1/4}|k|$ retains a cusp.

For fixed integer R\'enyi order $\alpha\geq2$, the singularity,
however, does not occur.  Here the R\'enyi CMI denotes the entropy combination
$S_\alpha(AB)+S_\alpha(BC)-S_\alpha(B)-S_\alpha(ABC)$.  To see this, note that the moments
$\Tr\rho^\alpha$ give the single-mode entropy
$h_\alpha(x)=\log[x^\alpha+(1-x)^\alpha]/(1-\alpha)$,
which is analytic at $x=0,1$.
The corresponding CMI combination of R\'enyi entropies therefore
decays exponentially with $|B|$ in both examples, including the
model in Eq.~\eqref{eq:model}.
Thus the power-law tails appear in the von Neumann CMI
but are absent from its integer-order R\'enyi counterparts.

\section{Discussion}
\label{sec:discussion}

We have shown that finite Markov length is unstable under local quantum dissipation. Starting from a product state, an arbitrarily short strictly local evolution can generate an algebraic CMI tail and hence infinite Markov length. The free-fermion model establishes this exactly at every finite time, and the auxiliary construction transfers it to a strictly local spin system. Using similar techniques, we have also constructed local Lindbladians rapidly relaxing to steady states with algebraic decay of CMI, even when arbitrarily weakly perturbing a product-state generator.  
% Infinite Markov length alone therefore implies neither slow parent dynamics nor an obstruction to dissipative local reversal evolution.

These examples are directly relevant to the infrared perspective developed in Ref.~\cite{ChenGroverRG2026}, which studies CMI in a slab geometry with macroscopic regions separated by a growing buffer.
In our states, $\lim_{L\to\infty}I(aL:cL\mid bL;t)=0$ at every fixed $t>0$ (Supplemental Sec.~\ref{app:infrared}). Their algebraic tail may therefore be viewed as a UV correction to an infrared description with zero CMI \cite{recent}. While this tail obstructs equivalence to a product state under superpolynomially accurate, gate-by-gate locally reversible channels, we have shown that these states are nonetheless connected to the initial product state via locally reversible circuits {\it with sublinear locality and inverse polynomial error}.  As also suggested in \cite{recent}, whether this locality and error tolerance in locally reversible circuits provides a fully satisfactory definition of phases should be carefully analyzed.  

Several simpler spin models raise questions beyond our solvable construction. For jump operators $ J_j =X_j+X_{j+1}$ acting on a product state, perturbation theory shows that the CMI has the same $tb^{-4}$ tail, while finite-size data at later times suggest a crossover to faster decay. It remains unclear whether this reflects a change in the asymptotic CMI at finite time or a finite-size effect (Supplemental Sec.~\ref{sec:response}). Resolving this would help establish whether simple one-dimensional spin systems can undergo sharp transitions in Markov length under local dissipation.

% Such a transition would also be relevant to the infrared perspective developed in Ref.~\cite{ChenGroverRG2026}, which studies CMI in a slab geometry with macroscopic regions separated by a growing buffer. Establishing a transition in the spin model above and examining its infrared CMI would help clarify whether a change in Markov length necessarily reflects a change in infrared phase structure, or can occur between states whose infrared CMI vanishes on both sides.

% Amirreza: does this make sense to add?
% In the infrared limit considered in
% Ref.~\cite{ChenGroverRG2026}, our one-dimensional evolutions
% satisfy $\lim_{L\to\infty}I(aL:cL\mid bL;t)=0$ at fixed finite
% $t>0$ (Supplemental Sec.~\ref{app:infrared}). Their algebraic
% CMI is therefore cutoff-suppressed~\cite{recent}. However,
% stacking $L^2$ independent copies of the Bogoliubov family,
% each contributing CMI proportional to $L^{-2}$, gives a
% three-dimensional state with finite, nonzero continuum CMI.
% This is genuine, continuum-surviving power-law
% CMI~\cite{recent}, yet these states arise from a product state
% under arbitrarily short local Lindblad evolution and return
% to it under onsite reset channels. Thus vanishing continuum
% CMI is unstable within the trivial phase and its survival alone does not
% diagnose phase nontriviality.

Another question is how this instability affects error correction and local recovery. Supplemental Sec.~\ref{app:surface-recovery} studies a planar surface code under correlated local noise generated by $Z_{r,c}+Z_{r,c+1}$, closely related to the spin noise above. At short times, the logical error improves with code size, while decoding inside a growing buffer region increasingly reproduces the global decoder's decisions. These motivate a direct study of the relation between CMI decay and the locality of quantum error correction, e.g. whether algebraic CMI decay allows for a sublinear locality of the decoder.

Finally, it is worth understanding which types of channels can diverge Markov length.  We have shown that certain coherent channels suffice, whereas certain incoherent channels \cite{ZhangGopalakrishnan2025Stability} cannot.  

{\it Note added}: During the completion of this draft, we noticed a closely related preprint \cite{veryrecent} which also provides examples of divergent Markov length under infinitesimal evolution.

% The stationary-state construction gives a complementary result. Strictly local, rapidly relaxing Lindbladians can have unique steady states with infinite Markov length, even under an arbitrarily weak perturbation of a product-state generator. 

% Together with the exact return along the original trajectory and the vanishing infrared CMI, these examples separate several notions that need not coincide: finite Markov length, local reversibility, rapid relaxation, two-way dissipative connectivity, and conditional information that survives at macroscopic scales. Which of these properties is needed to characterize mixed-state phases remains open.

\medskip
\medskip

\begin{acknowledgments}
We thank Yu-Hsueh Chen, Dominic Else, Raz Firanko, Tarun Grover, Zhi Li, Shengqi Sang for useful discussions.  We also acknowledge the use of AI in assisting with writing, simulations, and analytical derivations.  This research was supported by the Natural Sciences and Engineering Research Council of Canada (NSERC) and an Ontario Early Researcher Award. Research at Perimeter Institute is supported in part by the Government of Canada through the Department of Innovation, Science and Industry Canada and by the Province of Ontario through the Ministry of Colleges and Universities.
\end{acknowledgments}

% ===== Bibliography =====
\bibliographystyle{apsrev4-2}
\bibliography{references}

% ===== supplemental.tex =====
\onecolumngrid
\clearpage
\setcounter{section}{0}
\setcounter{subsection}{0}
\setcounter{equation}{0}
\setcounter{figure}{0}
\setcounter{table}{0}
\renewcommand{\thesection}{S\Roman{section}}
\renewcommand{\theequation}{S\arabic{equation}}
\renewcommand{\thefigure}{S\arabic{figure}}
\renewcommand{\thetable}{S\arabic{table}}
% Keep hyperlink destinations distinct from those in the main article.
\renewcommand{\theHsection}{supp.\arabic{section}}
\renewcommand{\theHsubsection}{supp.\arabic{section}.\arabic{subsection}}
\renewcommand{\theHequation}{supp.\arabic{equation}}
\renewcommand{\theHfigure}{supp.\arabic{figure}}
\renewcommand{\theHtable}{supp.\arabic{table}}
\begin{center}
 {\large\bfseries Supplemental Material\par}
 \vspace{0.4em}
 {\bfseries Instability of Markov chains under infinitesimal coherent dissipation\par}
 \vspace{0.5em}
 Amir-Reza Negari, Yijian Zou, and Timothy H. Hsieh
\end{center}
\vspace{0.5em}

The Supplemental Material follows the argument of the main text. We begin by explaining how local dissipation generates an infinite Markov length. Section~\ref{app:response} gives a perturbative explanation of the algebraic CMI at short times. Section~\ref{app:finite-time} then establishes an algebraic lower bound at every finite positive time, providing analytical support for the numerical results in Sec.~\ref{sec:fermions}. Section~\ref{app:spins} carries these results over to the local spin construction of Sec.~\ref{sec:spins}, detailing the auxiliary fermions and the Jordan--Wigner transformation that make the spin dynamics local. We then turn from evolving states to steady states and show that long-range CMI can coexist with rapid relaxation. Section~\ref{app:preparation} develops a solvable fermionic Gaussian example, and Sec.~\ref{app:spin-steady-parent} extends it to spins. Together, these sections support Sec.~\ref{sec:preparation} of the main text, including the rapid local convergence and local indistinguishability of the spin steady states. These constructions lead to the questions about mixed-state phases raised in Sec.~\ref{sec:implications}. Section~\ref{app:reverse-and-recovery} shows that the noise-driven trajectory can be retraced by an exponentially local evolution, while distinguishing this reversal from gate-by-gate local reversibility. Section~\ref{app:approx-local-reversibility} proves that superpolynomial CMI decay is preserved under gate-by-gate locally reversible evolution with a sublinear light cone and superpolynomially small total error. We then examine what the long-range correlations reveal: Sec.~\ref{app:correlations} shows that nonlinear correlations also become unstable, while Sec.~\ref{app:infrared} establishes that the CMI nevertheless vanishes in the infrared limit, including the slab geometry. Finally, we explore the open questions discussed in Sec.~\ref{sec:discussion}. Section~\ref{sec:response} studies simpler spin dynamics whose numerical behavior suggests a possible phase transition from algebraic CMI at short times to exponential decay at later times. Section~\ref{app:surface-recovery} examines local decoding in a surface code under correlated noise, testing whether effective local recovery can persist despite the instability of Markov length.

% Record only supplemental headings in this contents file.
\makeatletter
\let\articleaddtocontents\addtocontents
\renewcommand{\addtocontents}[2]{%
  \def\toc@extension{#1}\def\toc@main{toc}%
  \ifx\toc@extension\toc@main
    \articleaddtocontents{stoc}{#2}%
  \else
    \articleaddtocontents{#1}{#2}%
  \fi
}
\begingroup
  \let\addcontentsline\@gobblethree
  \section*{Contents}
\endgroup
\begingroup
  \small
  \hypersetup{linkcolor=black}
  \let\l@subsubsection\@gobbletwo
  \def\l@f@section{%
    \addpenalty{\@secpenalty}%
    \addvspace{0.5em plus\p@}%
  }
  \@starttoc{stoc}
\endgroup
\makeatother

\section{Short-time CMI in the fermion model}
\label{app:response}

We use the notation of the main text: $a,b,c$ are the lengths of
consecutive intervals $A,B,C$, and all logarithms are natural.
Thus entropies are measured in nats and
$I(A:C|B)=S_{a+b}+S_{b+c}-S_b-S_{a+b+c}$.
We derive the exact covariance and then compute the initial CMI
response. Section~\ref{app:finite-time} proves the separate result
needed at fixed positive time.

\subsection{Exact covariance and interval entropies}

For the jumps $F_j=c_j^\dagger+c_{j+1}^\dagger$, let
$K_{ij}=2\delta_{ij}+\delta_{i,j+1}+\delta_{i,j-1}$ on the infinite
chain. The anticommutation relations give for the covariance matrix
$\dot N=K-\{K,N\}/2$. Starting from the vacuum, the state remains
Gaussian, and
\begin{equation}
 N(t)=\id-e^{-tK},\qquad
 n(k,t)=1-e^{-t(2+2\cos k)}.
 \label{eq:exact-creation-covariance}
\end{equation}
For an interval, first compute this infinite-chain covariance and
then restrict it: $N_L=(N(t))_L$. In general this is different from
$\id_L-e^{-tK_L}$, because the interval exchanges particles with
its surroundings. If $n_\alpha$ are the eigenvalues of $N_L$, then
\begin{equation}
 S_L(t)=\sum_{\alpha=1}^L h(n_\alpha),\qquad
 h(n)=-n\log n-(1-n)\log(1-n).
 \label{eq:supp-gaussian-entropy}
\end{equation}
These formulas determine the finite-time CMI plotted in
Fig.~\ref{fig:gaussian} and provide a check at short times.

\subsection{The initial CMI response}
\label{app:fermion-response}

For the vacuum evolution in Sec.~\ref{sec:fermions}, expanding the
exact covariance gives $N_L(t)=tK_L+O_L(t^2)$ at fixed interval
length $L$. The eigenvalues of $K_L$ are
$\kappa_q=2+2\cos[\pi q/(L+1)]$, $q=1,\ldots,L$, so the
occupation eigenvalues are $t\kappa_q+O_L(t^2)$.
Using $h(n)=n(1-\log n)+O(n^2)$ and $\sum_q\kappa_q=2L$
gives
\begin{equation}
 S_L(t)=2Lt(1-\log t)+t h_L+O_L(t^2|\log t|),\qquad h_L=-\sum_{q=1}^L\kappa_q\log\kappa_q.
 \label{eq:fermion-response-entropy}
\end{equation}
The extensive term cancels from CMI, whose initial derivative is
therefore $\chi_{a,c}(b)=h_{a+b}+h_{b+c}-h_b-h_{a+b+c}$.
Sec.~\ref{app:response-asymptotics} evaluates $h_L$ and shows
that this combination is $3\zeta(3)ac/b^4+O(b^{-5})$ for fixed
$a,c$, giving the initial response stated in Sec.~\ref{sec:fermions}.
The short-time remainder is controlled at fixed interval sizes,
so this expansion alone does not establish a fixed-time CMI tail.

\subsection{The power-law tail of the initial response}
\label{app:response-asymptotics}

To evaluate $h_L$, we use the cosine expansion
\begin{equation}
 -(2+2\cos\theta)\log(2+2\cos\theta) =-2-3\cos\theta +4\sum_{m=2}^{\infty} \frac{(-1)^{m+1}\cos(m\theta)}{m(m^2-1)}.
 \label{eq:spin-fourier}
\end{equation}
This follows by multiplying the cosine series for
$\log(2+2\cos\theta)$ by $-(2+2\cos\theta)$; the product is
continuous at $\theta=\pi$, with $0\log0=0$.

On the grid $\theta_q=\pi q/(L+1)$, odd harmonics sum to zero.
Even harmonics sum to $-1$, except when their index is a multiple of
$2(L+1)$, in which case the sum is $L$. Substitution in
Eq.~\eqref{eq:spin-fourier} gives the convergent expression
\begin{equation}
 h_L=-2(L+1)+\log16
       -2\sum_{p=1}^{\infty}
          \frac{1}{p[4(L+1)^2p^2-1]}.
 \label{eq:spin-h-exact}
\end{equation}
Expanding the denominator in inverse powers of $L+1$ yields
\begin{equation}
 h_L=-2(L+1)+\log16 -\frac{\zeta(3)}{2(L+1)^2} -\frac{\zeta(5)}{8(L+1)^4}+O(L^{-6}).
 \label{eq:spin-h-asymptotic}
\end{equation}
The terms linear and constant in $L$ cancel from the four-entropy
combination. The leading contribution is therefore
\begin{equation}
 \chi_{a,c}(b)=\frac{\zeta(3)}2 \biggl[\frac1{(b+1)^2}+\frac1{(a+b+c+1)^2} -\frac1{(a+b+1)^2}-\frac1{(b+c+1)^2}\biggr] +O_{a,c}(b^{-6}).
 \label{eq:spin-chi-bracket}
\end{equation}
Here the explicit $(L+1)^{-4}$ term in
Eq.~\eqref{eq:spin-h-asymptotic} contributes only at order $b^{-6}$
for fixed $a,c$. Expanding the bracket gives
\begin{equation}
 \chi_{a,c}(b)=\frac{3\zeta(3)ac}{b^4}+O_{a,c}(b^{-5}).
 \label{eq:response}
\end{equation}
This is the large-buffer behavior of the derivative at
$t=0^+$; it does not determine the large-buffer behavior at a fixed
positive time.

\section{A power-law CMI lower bound for fermions at positive times}
\label{app:finite-time}

We prove that the evolution has infinite Markov length at every
fixed positive time, using the actual restricted covariance rather
than a function of a finite-interval Laplacian.  We establish a positive
lower bound proportional to $b^{-4}$ by writing the CMI as a positive
integral and retaining paths that propagate across the buffer.

Write $K=2I-T-T^\dagger$ on the infinite chain and set
$Q_t=e^{-tK}$.  The covariance is $N_t=I-Q_t$.  For a finite set $R$,
its restriction is therefore $N_{t,R}=I_R-(Q_t)_R$, where the subscript
denotes a principal submatrix.  

\subsection{Entropy as a positive integral}

For $0\leq x\leq1$, the single-mode fermionic entropy obeys
\begin{equation}
 h(x)=\int_0^\infty du\,
 \log\frac{(x+u)(1-x+u)}{u(1+u)}.
 \label{eq:creation-entropy-integral}
\end{equation}
For a positive matrix $M$ indexed by disjoint sets $A,B,C$, define
\begin{equation}
 J(M)=\log\frac{\det M_{AB}\det M_{BC}}
                       {\det M_B\det M_{ABC}}.
 \label{eq:creation-logdet-cmi}
\end{equation}
This quantity is nonnegative by the determinant form of strong
subadditivity.  Applying Eq.~\eqref{eq:creation-entropy-integral} to each
restricted covariance gives
\begin{equation}
 I(A:C|B)=\int_0^\infty du\,
 \bigl[J(N_t+uI)+J(I-N_t+uI)\bigr].
 \label{eq:creation-cmi-positive-integral}
\end{equation}
Only the indicated principal submatrices enter $J$; no entropy of an
infinite region is taken.  Each term in the integral is nonnegative,
so retaining its first term over any interval of $u$ gives a lower bound.

First take $A$ and $C$ to be individual sites with $b$ sites between
them.  Put $M=N_t+uI$ and eliminate the buffer $B$ by its Schur
complement.  The resulting two-by-two matrix has diagonal entries
$d_A,d_C>0$ and off-diagonal entry
\begin{equation}
 s=M_{AC}-M_{AB}M_B^{-1}M_{BC}.
 \label{eq:creation-schur-cross}
\end{equation}
Equation~\eqref{eq:creation-logdet-cmi} then becomes
$J(M)=-\log[1-|s|^2/(d_Ad_C)]\geq |s|^2/(d_Ad_C)$.
All off-diagonal entries of $M$ are nonpositive, whereas
\begin{equation}
 M_B^{-1}
 =\sum_{n=0}^\infty\frac{[(Q_t)_B]^n}{(1+u)^{n+1}}
 \label{eq:creation-positive-resolvent}
\end{equation}
has nonnegative entries.  Thus the direct and buffer-mediated terms in
Eq.~\eqref{eq:creation-schur-cross} have the same sign.
Keeping only the nearest-neighbor connection at either end yields
\begin{equation}
 |s|\geq t^2e^{-4t}(M_B^{-1})_{1b}.
 \label{eq:creation-schur-lower}
\end{equation}
Also $d_A,d_C\leq2t+u$: the Schur complement can only decrease a
diagonal entry, and $I-e^{-tK}\leq tK$ implies $(N_t)_{jj}\leq2t$.

\subsection{Propagation through the buffer}

Let $K_b$ be the Dirichlet Laplacian on the intervening interval, with
diagonal two and nearest-neighbor entries minus one.  The unrestricted
kernel counts every random-walk path that remains inside this interval.  Therefore
$(Q_t)_B\geq e^{-tK_b}$ entry by entry.  Products preserve this
inequality because all entries are nonnegative.  Inserting it into
Eq.~\eqref{eq:creation-positive-resolvent} gives the exact comparison
\begin{equation}
 (M_B^{-1})_{1b}\geq
 \sum_{n=0}^\infty
 \frac{(e^{-ntK_b})_{1b}}{(1+u)^{n+1}}.
 \label{eq:creation-killed-comparison}
\end{equation}

Set $m=b+1$.  The Dirichlet heat kernel satisfies, for a numerical
constant $c_0>0$ independent of $b$,
\begin{equation}
 (e^{-\tau K_b})_{1b}\geq\frac{c_0}{m^3},
 \qquad m^2\leq\tau\leq2m^2.
 \label{eq:creation-heat-endpoints}
\end{equation}
Its sine expansion is
\begin{equation}
 (e^{-\tau K_b})_{1b}
 =\frac{2}{m}\sum_{j=1}^{m-1}(-1)^{j+1}
 \sin^2\frac{\pi j}{m}
 \exp\left[-4\tau\sin^2\frac{\pi j}{2m}\right].
 \label{eq:creation-heat-sine-sum}
\end{equation}
The first term controls this sum in the time interval of
Eq.~\eqref{eq:creation-heat-endpoints}: the absolute remainder
relative to that term is bounded by
$\frac{\pi^2}{4}\sum_{j\geq2}j^2e^{-(4j^2-\pi^2)}<\frac12$,
using $\sin(\pi/m)\geq2/m$,
$4\sin^2(\pi j/2m)\geq4j^2/m^2$, and
$4\sin^2(\pi/2m)\leq\pi^2/m^2$.
Thus, for example, $c_0=4e^{-2\pi^2}$ is sufficient.
The series bound follows from
$\sum_{j\geq2}j^2e^{-4j^2}<5e^{-16}$ and $\pi^2<10$.

For $m^2\geq2t$, there are at least $m^2/(2t)$ integers $n$ with
$m^2\leq nt\leq2m^2$.  If $0<u\leq t/m^2$, their weights in
Eq.~\eqref{eq:creation-killed-comparison} are at least $e^{-3}$.
Consequently
\begin{equation}
 (M_B^{-1})_{1b}\geq\frac{c_1}{tm},
 \qquad 0<u\leq\frac{t}{m^2},
 \qquad c_1=\frac{c_0e^{-3}}2.
 \label{eq:creation-resolvent-lower}
\end{equation}
Combining this with $d_A,d_C\leq3t$ and
Eq.~\eqref{eq:creation-schur-lower} gives
$J(N_t+uI)\geq c_1^2e^{-8t}/(9m^2)$ throughout this interval.
Integration therefore proves
\begin{equation}
 I(A:C|B;t)\geq
 \frac{c_1^2}{9}\frac{te^{-8t}}{(b+1)^4},
 \qquad (b+1)^2\geq2t.
 \label{eq:creation-fixed-time-cmi-lower}
\end{equation}
The small numerical prefactor is not intended to estimate the actual
CMI.  It establishes an algebraic lower bound at every fixed $t>0$.  Enlarging
either endpoint interval can only increase the CMI, so the result also
applies to longer $A$ and $C$.

\subsection{Buffers on both sides of a region}

The same argument covers the geometry used to define Markov length.
Take $A$ to be one site, put $b$ buffer sites on each side, and retain
one site of $C$ just beyond the right buffer.  Now $M_B$ includes both
buffer components.  Its nonnegative resolvent expansion includes all
the paths confined to the right component, so its relevant endpoint
entry still obeys Eq.~\eqref{eq:creation-resolvent-lower}.
The remaining Schur-complement bounds are unchanged.  Thus
Eq.~\eqref{eq:creation-fixed-time-cmi-lower} also holds with a
two-sided buffer.  Enlarging $C$ to the entire exterior only increases
the CMI.  We can work throughout on finite open chains whose length
is proportional to $b$, then take the thermodynamic limit.
An algebraic lower bound is incompatible with exponential decay even
when the latter is multiplied by a fixed polynomial in region sizes.

For the auxiliary-spin construction, the entropy identity is needed
only for contiguous intervals.  On a finite open chain choose $A$ at
the left endpoint, $B$ as the next $b$ complete cells, and $C$ as the
remaining cells.  Then $B$ shields $A$ from the rest of the chain and
all four regions entering the CMI are contiguous.  The argument above remains valid, uniformly in the chain length, and the
spin and fermionic CMIs agree exactly.  This establishes the same
absence of a size-independent finite Markov length for the local open
spin chain, with physical spins at $2j$ and auxiliary spins at $2j+1$.
No identity for disconnected spin marginals is assumed here.
Neither this lower bound nor its spin realization establishes the
sharper asymptotic coefficient $3\zeta(3)tac$; that coefficient remains
the fixed-time prediction supported by Fig.~\ref{fig:gaussian}.

% \subsection{Bosonic comparison}
% \label{app:bosons}

% For this comparison, return to the main-text phase convention,
% $K=2\id+T+T^\dagger$, whose symbol is $\kappa(k)=2+2\cos k$.
% The bosonic counterpart has jumps
% $a_j^\dagger+a_{j+1}^\dagger$ and starts from the vacuum. Its initial CMI
% response is again Eq.~\eqref{eq:response}, and its evolution is also
% Gaussian. The occupations now obey $\dot n_{\rm b}=\kappa(1+n_{\rm b})$,
% so $N_{\rm b}(t)=e^{tK}-\id$; the entropy follows from
% $h_{\rm b}(n)=(1+n)\log(1+n)-n\log n$. Near $k=\pi$,
% $n_{\rm b}(\pi+q,t)=tq^2+O_t(q^4)$, and the bosonic data in
% Fig.~\ref{fig:gaussian-comparison} support the same fixed-time algebraic tail.
% The rigorous bound above is for fermions; for bosons a similar analytical proof can be established.

% \begin{figure}[tbp]
%  \centering
%  \includegraphics[width=0.45\textwidth]{Figures/gaussian_cmi_beyond_perturbative_PRX.pdf}
%  \caption{Comparison of the fermionic and bosonic Gaussian evolutions.
%  CMI divided by time is shown for one-site end regions at
%  $t=10^{-2},10^{-1},1,10$. Filled circles denote fermions and open
%  diamonds bosons. The buffer length is $B$, and the dotted guide is
%  $3\zeta(3)B^{-4}$. The bosonic data support the same fixed-time
%  algebraic tail.}
%  \label{fig:gaussian-comparison}
% \end{figure}

\section{Mapping the fermionic evolution to a local spin model}
\label{app:spins}

\subsection{Adding auxiliaries without changing the physical evolution}

Use physical Majoranas $\eta_j=c_j+c_j^\dagger$ and
$\xi_j=-i(c_j-c_j^\dagger)$, and auxiliary Majoranas
$\gamma_j,\delta_j$. Every Majorana squares to one; any two
distinct Majoranas anticommute, including physical and auxiliary
ones. In particular, $\gamma_jF_j=-F_j\gamma_j$ and
$\gamma_jF_j^\dagger=-F_j^\dagger\gamma_j$.

Write the physical density matrix as an even polynomial
$r(\eta,\xi)$, normalized in its $M$-mode space. In the enlarged
algebra the same polynomial has its spectrum repeated $2^M$ times,
so the normalized state is $\widehat\rho=2^{-M}r(\eta,\xi)$.
It contains no auxiliary Majoranas. This is the maximally mixed
auxiliary state, specified directly in the Majorana algebra.

For $\widetilde F_j=i\gamma_jF_j$, the operator identities are
\begin{equation}
 \widetilde F_j\widehat\rho\,\widetilde F_j^\dagger =2^{-M}\gamma_jF_jrF_j^\dagger\gamma_j =2^{-M}F_jrF_j^\dagger,\qquad \widetilde F_j^\dagger\widetilde F_j=F_j^\dagger F_j.
 \label{eq:lift-identity}
\end{equation}
In the first identity, $\gamma_j$ commutes with $r$ and anticommutes
with both odd jumps. Moving it to the right gives two minus signs
and then $\gamma_j^2=1$. Thus the lifted dissipator reproduces
the original one, and the state remains an even polynomial of
physical Majoranas throughout the evolution.

\subsection{The local spin jumps}

Order the spins as $0,1,2,3,\ldots$, with the physical spins at
$2j$ and auxiliary spins at $2j+1$. Put
$P_j=\prod_{\ell<j}Z_{2\ell}Z_{2\ell+1}$. The Jordan--Wigner
representation is
\begin{equation}
 \eta_j=P_jX_{2j},\qquad \xi_j=P_jY_{2j},\qquad \gamma_j=P_jZ_{2j}X_{2j+1},\qquad \delta_j=P_jZ_{2j}Y_{2j+1}.
\end{equation}
Direct multiplication gives
$\gamma_jc_j^\dagger=-\sigma^-_{2j}X_{2j+1}$ and
$\gamma_jc_{j+1}^\dagger=-iY_{2j+1}\sigma^-_{2j+2}$.
Removing the irrelevant common phase gives Eq.~\eqref{eq:spin-jump}.
The initial state is $4^{-M}\prod_j(1-i\eta_j\xi_j)
=4^{-M}\prod_j(1+Z_{2j})$.

The formulas apply directly on the infinite chain and on an open
chain with internal bond jumps. A closing bond on a fermionic ring
requires separate parity conventions, so no periodic local-spin
identification is used here.

\subsection{Interval entropies and open boundaries}

For a consecutive interval of $L$ complete cells, the string to
the left of the interval cancels in the even physical Majorana
polynomial. Its spin spectrum therefore consists of every fermion
eigenvalue divided by $2^L$ and repeated $2^L$ times. This proves
the entropy identity in Sec.~\ref{sec:spins} and hence the CMI
equality in Eq.~\eqref{eq:spin-cmi} for consecutive $A,B,C$. This does
not assert an identity for arbitrary disconnected spin marginals.

For a boundary region on a finite open chain, $A$ can be the first
cell, $B$ the next $b$ cells, and $C$ the remainder. All four
entropy regions are contiguous. The uniform heat-kernel lower bound
in Sec.~\ref{app:finite-time} then excludes a system-size-independent
finite Markov length in this shielding geometry as well.

\section{Fermionic steady states with power-law CMI}
\label{app:preparation}
\subsection{Exact evolution and steady state}
\label{app:bogoliubov-parent}
\label{app:grassmann-steady-state}

We solve the evolution generated by
Eq.~\eqref{eq:bogoliubov-lindbladian}, then restrict the resulting
state to an interval. We consider parity-even fermionic states,
including non-Gaussian inputs. Use the ordered Majoranas
$w=(\eta_0,\xi_0,\eta_1,\xi_1,\ldots)$ and the convention
\begin{equation}
 \Gamma_{ab}=\frac{i}{2}\Tr\rho[w_a,w_b].
 \label{eq:majorana-covariance-convention}
\end{equation}
Thus a vacuum mode has $\Gamma_{\eta,\xi}=-1$.
The cooling modes are $c_j=(\eta_j+i\xi_j)/2$ and
$d_j=(\eta_j+i\xi_{j+1})/2$, with rates $1-p$ and $p$.
Define the antisymmetric matrix $\Gamma_p$ by
\begin{equation}
 (\Gamma_p)_{\eta_j,\xi_\ell}
 =-(1-p)\delta_{j\ell}-p\delta_{\ell,j+1},
 \label{eq:supp-bogoliubov-real-covariance}
\end{equation}
with all other independent entries zero. We will find that this
is the stationary covariance.

A generating function keeps track of all even correlations at once.
Introduce Grassmann variables
$\theta_a\theta_b=-\theta_b\theta_a$ and $\theta_a^2=0$.

For an ordered finite set $I={i_1<\cdots<i_{2m}}$, put
\begin{equation}
C_I(\rho)=i^m\Tr(\rho w_{i_1}\cdots w_{i_{2m}}),\;\;\
\chi_\rho(\theta)=\sum_{|I|\ {\rm even}}
C_I(\rho)\theta_{i_1}\cdots\theta_{i_{|I|}},
\label{eq:grassmann-definition}
\end{equation}
with $C_\varnothing=1$. For an even state these coefficients
determine all local reduced density matrices. In particular,
$C_{{a,b}}=\Gamma_{ab}$ for $a<b$. For an even Gaussian state,
the Grassmann generating function is
\begin{equation}
\chi_\rho(\theta)=
\exp\left(\frac12\theta^{\mathsf T}\Gamma\theta\right).
\label{eq:grassmann-gaussian}
\end{equation}

For the full-chain generator $\mathcal L_p$, the canonical
anticommutation relations give the hierarchy
\begin{equation}
\partial_s C_I=-m C_I+
\sum_{1\leq r<u\leq2m}(-1)^{r+u-3}
(\Gamma_p)_{i_r i_u}C_{I\setminus{i_r,i_u}}.
\label{eq:grassmann-moment-hierarchy}
\end{equation}
This identity follows directly from
$\mathcal L_p^\dagger(O)=\frac12\sum_F
[F^\dagger[O,F]+[F^\dagger,O]F]$.
For an even Majorana product, each commutator removes one of its
Majoranas. Multiplication by the remaining jump either restores
that Majorana, giving $-1/2$ per factor, or removes a second one,
giving the displayed contraction. The sign comes from bringing
positions $r,u$ to the front. For $|I|=2$, the identity reduces to
$\partial_s\Gamma=\Gamma_p-\Gamma$, fixing the sign convention.

Multiplying the hierarchy by the corresponding Grassmann monomials
gives
\begin{equation}
\partial_s\chi_s=
\left[-\frac12\sum_a\theta_a\partial_{\theta_a}
+\frac12\theta^{\mathsf T}\Gamma_p\theta\right]\chi_s.
\label{eq:grassmann-evolution}
\end{equation}
Here derivatives are left Grassmann derivatives; the first term
counts the degree of each monomial. Its solution is
\begin{equation}
\chi_s(\theta)=
\exp\left[\frac{1-e^{-s}}2\theta^{\mathsf T}\Gamma_p\theta\right]
\chi_0(e^{-s/2}\theta).
\label{eq:grassmann-solution}
\end{equation}
Differentiation verifies the formula and its initial condition.
As $s\to\infty$, every nonconstant contribution from the initial
state vanishes. The limiting state is therefore the unique even
stationary state and is Gaussian with covariance $\Gamma_p$; we denote it
by $\rho_\infty(p)$. Taking the quadratic terms in
Eq.~\eqref{eq:grassmann-solution} gives, for any initial state,
\begin{equation}
\Gamma(s)=\Gamma_p+e^{-s}[\Gamma(0)-\Gamma_p].
\label{eq:bogoliubov-covariance-relaxation}
\end{equation}
With the Fourier convention
$\Gamma_p(k)=\sum_r(\Gamma_p)_{0,r}e^{ikr}$, the stationary
covariance is equivalently
\begin{equation}
 \Gamma_p(k)=
 \begin{pmatrix}0&-q_p(k)\\q_p(k)^*&0\end{pmatrix},
 \qquad q_p(k)=1-p+pe^{ik}.
 \label{eq:bogoliubov-covariance}
\end{equation}

Let $R$ be any interval of $M$ sites, let $\rho_R(s)$ denote the reduced density
matrix of $e^{s\mathcal L_p}(\rho_0)$ on $R$, and let
$\rho_{\infty,p,R}$ be the restriction of $\rho_\infty(p)$ to the
same interval. Setting all Grassmann variables outside $R$ to zero
in Eq.~\eqref{eq:grassmann-solution} gives
\begin{equation}
\chi_{\rho_R(s)}(\theta_R)=
\exp\left[
\frac{1-e^{-s}}2\theta_R^{\mathsf T}\Gamma_{p,R}\theta_R
\right]
\chi_{\rho_{0,R}}(e^{-s/2}\theta_R),
\label{eq:grassmann-restricted-solution}
\end{equation}
where $\Gamma_{p,R}$ is the principal submatrix of $\Gamma_p$ on $R$.
Thus the reduced state is obtained by first evolving the full
infinite chain and only then restricting to $R$.

We now derive a trace-norm bound directly from this formula.
Write $\sigma=\rho_{\infty,p,R}$ and introduce an auxiliary system
$E$ of $M$ fermionic modes in the same state $\sigma$.
We will mix the two systems by a Gaussian
unitary~\cite{Bravyi2005FLO}, whose action we give explicitly.
Let $w_a$ and $v_a$, $a=1,\ldots,2M$, be the subsystem and
auxiliary Majoranas. Each family satisfies the Majorana
anticommutation relations, and every $w_a$ anticommutes with
every $v_b$.
On the ordinary tensor product this means
$w_a=w_a^{(R)}\otimes\id$ and
$v_a=P_R\otimes w_a^{(E)}$, where $P_R$ is the subsystem
fermion parity. Thus $\tau\otimes\sigma$ is a well-defined product
state for any even subsystem density matrix $\tau$.

Consider the Hermitian operator and unitary
\begin{equation}
 H=\frac{i}{2}\sum_{a=1}^{2M}w_av_a,\qquad
 U_\alpha=e^{-i\alpha H},\qquad 0\leq\alpha\leq\frac{\pi}{2}.
 \label{eq:grassmann-mixing-unitary}
\end{equation}
The commutators $[H,w_a]=-iv_a$ and $[H,v_a]=iw_a$ give
\begin{equation}
 U_\alpha^\dagger w_aU_\alpha
 =(\cos\alpha)w_a+(\sin\alpha)v_a.
 \label{eq:grassmann-majorana-rotation}
\end{equation}
After tracing out $E$, this defines the channel
$\Phi_\alpha(\tau)=\Tr_E[U_\alpha(\tau\otimes\sigma)U_\alpha^\dagger]$.
Because $U_\alpha$ preserves total parity, $\Phi_\alpha(\tau)$
is even whenever $\tau$ is even.
Its generating function is
\begin{equation}
 \chi_{\Phi_\alpha(\tau)}(\theta_R)
 =\chi_\tau(\cos\alpha\,\theta_R)
  \chi_\sigma(\sin\alpha\,\theta_R).
 \label{eq:grassmann-channel-factorization}
\end{equation}
To see this, expand each even Majorana product using
Eq.~\eqref{eq:grassmann-majorana-rotation}. A term with an odd
number of Majoranas from either system has zero expectation,
since both input states are even. In every surviving term the
subsystem and auxiliary expectations factorize. Moving the
subsystem Majoranas before the auxiliary ones gives exactly the
signs produced by multiplying the two Grassmann polynomials.
This proves Eq.~\eqref{eq:grassmann-channel-factorization}
coefficient by coefficient, for arbitrary even $\tau$.

Since $\sigma$ is Gaussian,
$\chi_\sigma(\theta_R)=
\exp(\theta_R^{\mathsf T}\Gamma_{p,R}\theta_R/2)$.
Choose the angle $\alpha(s)$ by $\cos\alpha(s)=e^{-s/2}$.
Equation~\eqref{eq:grassmann-channel-factorization} then becomes
exactly Eq.~\eqref{eq:grassmann-restricted-solution}.
Even Majorana monomials form a basis of the even operator algebra,
so equality of these generating functions determines the entire
reduced density matrix. Therefore
\begin{equation}
 \rho_R(s)=\Phi_{\alpha(s)}(\rho_{0,R}),\qquad
 \Phi_{\pi/2}(\tau)=\sigma.
 \label{eq:grassmann-state-identification}
\end{equation}
The second identity follows by setting $\cos\alpha=0$ in
Eq.~\eqref{eq:grassmann-channel-factorization} and using
$\chi_\tau(0)=1$.

It remains to compare these two channels. For any density matrix
$\Omega$ and unitaries $U,V$, adding and subtracting $V\Omega U^\dagger$
gives
\begin{equation}
 \|U\Omega U^\dagger-V\Omega V^\dagger\|_1 \leq\|(U-V)\Omega U^\dagger\|_1 +\|V\Omega(U^\dagger-V^\dagger)\|_1 \leq2\|U-V\|_\infty.
 \label{eq:grassmann-unitary-trace-bound}
\end{equation}
Here we used $\|AXB\|_1\leq\|A\|_\infty\|X\|_1\|B\|_\infty$
and $\|\Omega\|_1=1$.
Each of the $2M$ terms in $H$ has norm $1/2$, so
$\|H\|_\infty\leq M$. Integrating
$\partial_\alpha U_\alpha=-iHU_\alpha$ therefore gives
\begin{equation}
 \|U_{\alpha(s)}-U_{\pi/2}\|_\infty
 \leq M\left(\frac{\pi}{2}-\alpha(s)\right)
 =M\arcsin(e^{-s/2}).
 \label{eq:grassmann-angle-bound}
\end{equation}
Apply Eq.~\eqref{eq:grassmann-unitary-trace-bound} to
$\Omega=\rho_{0,R}\otimes\sigma$ and use trace-norm
contractivity under the partial trace. Together with
Eq.~\eqref{eq:grassmann-state-identification}, this yields
\begin{equation}
 \|\rho_R(s)-\rho_{\infty,p,R}\|_1
 \leq2M\arcsin(e^{-s/2})
 \leq\pi M e^{-s/2},
 \label{eq:bogoliubov-mixing}
\end{equation}
where $\arcsin x\leq\pi x/2$ for $0\leq x\leq1$.
The bound is uniform over parity-even initial states of the full
chain, including non-Gaussian states, and over $p\in[0,1]$.
For $0<\varepsilon\leq1$, it suffices to take
$s\geq2\log(\pi M/\varepsilon)$ to make the local trace-norm
error at most $\varepsilon$.

\subsection{Power-law CMI of the steady states}
\label{app:bogoliubov-cmi}

We compute the CMI of the stationary state
$\rho_\infty=\rho_\infty(p)$ for $0<p<1$.
For single-site end regions separated by $b\geq1$ sites, write
$I_b=I(A:C|B)_{\rho_\infty}$. We will prove the two-sided bound
\begin{equation}
 \frac{3\zeta(3)p(1-p)}{4(b+3)^4}
 \leq I_b\leq
 \frac{3\zeta(3)p(1-p)}{4(b-\tfrac12)^4}.
 \label{eq:bogoliubov-cmi-explicit}
\end{equation}
This gives a positive algebraic lower bound and the leading
coefficient stated in the main text. We prove it by evaluating an
interval determinant and then bounding a positive integral.

\paragraph{The interval determinant.}
Order the Majorana indices as
$\xi_0,\eta_0,\xi_1,\eta_1,\ldots$.
A diagonal change of phases turns $i\Gamma_p$ into $-W_p$,
where $W_p$ is a real nearest-neighbor matrix with alternating
bonds $1-p,p,1-p,p,\ldots$.
For an interval of $L$ physical sites, define
$E_L=(\id-W_{p,L})/2$, with $W_{p,L}$ the principal
$2L\times2L$ submatrix. Its Gaussian entropy is
$S_L=\tfrac12\Tr h(E_L)$, where
$h(x)=-x\log x-(1-x)\log(1-x)$.
Alternating signs on the Majorana indices send $W_{p,L}$ to
$-W_{p,L}$, so $E_L$ and $\id-E_L$ have the same spectrum.
In the entropy integral
\eqref{eq:creation-entropy-integral}, the two terms therefore
cancel the factor $1/2$, giving
\begin{equation}
 I_b=\int_0^\infty J_b(v)\,dv,\qquad J_b(v)=2\log D_{b+1}(v)-\log D_b(v)-\log D_{b+2}(v),
 \label{eq:bogoliubov-entropy-integral}
\end{equation}
where $D_L(v)=\det(E_L+v\id)$ and $D_0(v)=1$.

Put $\alpha=(1-p)/2$, $\beta=p/2$, and $d=1/2+v$.
The tridiagonal determinant recurrence, grouped in pairs of
Majorana indices, is
\begin{equation}
 D_1=d^2-\alpha^2,\qquad D_L=(d^2-\alpha^2-\beta^2)D_{L-1} -\alpha^2\beta^2D_{L-2}\qquad(L\geq2).
\end{equation}
Define $q=p(1-p)$, $r=p/(1-p)$, and $\theta>0$ by
$\cosh\theta=1+2v(1+v)/q$. Solving the recurrence gives
\begin{equation}
 D_L(v)=(\alpha\beta)^L
 \frac{\sinh[(L+1)\theta]+r\sinh(L\theta)}{\sinh\theta}.
 \label{eq:bogoliubov-determinant}
\end{equation}
Equivalently, the only part of $\log D_L$ that is not constant
or linear in $L$ is
$\log(1-R e^{-2L\theta})$, where
\begin{equation}
 R=\frac{e^{-\theta}+r}{e^\theta+r},
 \qquad e^{-2\theta}<R<1.
\end{equation}
All other terms cancel in $J_b$.

\paragraph{A positive integral and matching bounds.}
Let $g(x)=e^{-x}/(1-e^{-x})^2$ for $x>0$.
The function $\log(1-e^{-x})$ has second derivative $-g(x)$.
Writing its second finite difference as a double integral gives
the exact identity
\begin{equation}
 J_b(v(\theta))=
 \int_0^{2\theta}\!ds\int_0^{2\theta}\!dt\,
 g(2b\theta-\log R+s+t).
 \label{eq:bogoliubov-logdet-exact}
\end{equation}
Since $g$ is positive and decreasing, and
$0<-\log R<2\theta$, we obtain
\begin{equation}
 4\theta^2 g(2(b+3)\theta)
 \leq J_b(v(\theta))\leq
 4\theta^2 g(2b\theta).
 \label{eq:bogoliubov-integrand-bounds}
\end{equation}
The change of variables from $v$ to $\theta$ obeys
\begin{equation}
 \frac{dv}{d\theta}
 =\frac{q\sinh\theta}
 {2\sqrt{1+2q(\cosh\theta-1)}},
 \qquad
 \frac{q\theta}{2}\leq\frac{dv}{d\theta}
 \leq\frac{q\theta e^\theta}{2}.
\end{equation}
For the lower inequality, use $q\leq1/4$ to bound the square
root by $\cosh(\theta/2)$, then use
$\sinh(\theta/2)\geq\theta/2$.
For the upper inequality, bound the square root below by one
and use $\sinh\theta\leq\theta e^\theta$.
Finally, $e^\theta g(2b\theta)\leq g((2b-1)\theta)$ for
$b\geq1$. Combining these inequalities yields
\begin{equation}
 2q\int_0^\infty\theta^3g(2(b+3)\theta)\,d\theta \leq I_b \leq2q\int_0^\infty\theta^3g(2(b-\tfrac12)\theta)\,d\theta.
\end{equation}
For both integrals, we use the positive series
$g(x)=\sum_{n\geq1}ne^{-nx}$, which gives
\begin{equation}
 \int_0^\infty x^3g(x)\,dx=6\zeta(3).
\end{equation}
Rescaling $x$ proves Eq.~\eqref{eq:bogoliubov-cmi-explicit}.
In particular, the same bounds imply
\begin{equation}
 I(A:C|B)_{\rho_\infty(p)}
 =\frac{3\zeta(3)p(1-p)}{4b^4}
  +O\!\left(\frac{p(1-p)}{b^5}\right),
 \label{eq:bogoliubov-general-asymptotic}
\end{equation}
with an absolute remainder constant. Thus no expansion under
the entropy integral is needed.

The lower bound also holds when either end region is enlarged,
because discarding sites from $A$ or $C$ cannot increase CMI.
It applies to the stationary interval marginals found in
Sec.~\ref{app:bogoliubov-parent}. Placing $A$ at
an open-chain endpoint makes $B$ a shielding region, so the
same conclusion holds for the spin realization on complete
cells. At $p=0$ the state has only onsite pairing, and at $p=1$
it has only nearest-neighbor bond pairing; in both cases
$I_b=0$ for $b\geq1$.

\section{Auxiliary fermions and spin steady states}
\label{app:rapid-preparation}
\label{app:spin-steady-parent}
The auxiliary noise rapidly makes the auxiliary fermions locally
maximally mixed and removes their local correlations with the
physical fermions. After this, the lifted cooling terms
reproduce the Gaussian evolution studied in
Sec.~\ref{app:bogoliubov-parent}. We therefore use a triangle
inequality to bound the total relaxation error by the sum of the
auxiliary averaging error and the physical relaxation error.

\subsection{Exact auxiliary evolution}

As in Sec.~\ref{app:spins}, introduce an auxiliary fermion
$g_j=(\gamma_j+i\delta_j)/2$ at each physical site. We first work
on a periodic fermionic chain of $N\geq2$ cells, with all indices
modulo $N$. The auxiliary noise is
\begin{equation}
 \mathcal A_g=\sum_j\left(\D{Q_j^z}+\D{Q_j^b}\right),
 \qquad Q_j^z=-i\gamma_j\delta_j,\qquad
 Q_j^b=i\gamma_j\gamma_{j+1}.
 \label{eq:auxiliary-generator}
\end{equation}
Each jump is a Hermitian unitary, so $\D{Q}(X)=QXQ-X$.
Put $\mathcal E_Q(X)=(X+QXQ)/2$. The conjugation maps commute,
and the exact evolution is
\begin{equation}
 e^{s\mathcal A_g}
 =\prod_Q\left[\mathcal E_Q+
 e^{-2s}(\mathrm{id}-\mathcal E_Q)\right],
 \qquad \mathcal E_g=\prod_Q\mathcal E_Q.
 \label{eq:auxiliary-exact-channel}
\end{equation}
Each factor differs from its limiting average by at most
$e^{-2s}$ in induced trace norm. Replacing $K$ factors by their
averages therefore gives an error at most $K e^{-2s}$.

The onsite jumps remove occupation coherences, and the pair
flips connect all configurations of the same parity. Thus, for
an auxiliary density matrix $\tau$,
\begin{equation}
 \mathcal E_g(\tau)
 =2^{-N}\bigl[\id+\Tr(P_g\tau)P_g\bigr],
 \qquad P_g=\prod_j(1-2g_j^\dagger g_j).
 \label{eq:spin-auxiliary-parity-limit}
\end{equation}
Tracing out any auxiliary mode removes the parity term.
Every proper auxiliary interval therefore relaxes to the
normalized identity.

\subsection{Local relaxation and indistinguishable steady states}

Using the physical cooling modes of Sec.~\ref{app:bogoliubov-parent},
define
\begin{equation}
 \widetilde{\mathcal G}_p=\mathcal A_g+\mathcal B_p,\qquad
 \mathcal B_p=(1-p)\sum_j\D{i\gamma_jc_j}
                     +p\sum_j\D{i\gamma_jd_j}.
 \label{eq:spin-fermionic-bulk-parent}
\end{equation}
We consider arbitrary parity-even initial states. Every jump
preserves the total parity $P=P_fP_g$, where
$P_f=\prod_j(1-2c_j^\dagger c_j)$.
Auxiliary conjugation changes each cooling jump only by a sign,
so $\mathcal A_g$, $\mathcal E_g$, and $\mathcal B_p$ commute.

Let $R$ be an interval of $M<N$ complete cells. Only its $M$
onsite auxiliary jumps, $M-1$ internal bond jumps, and two
crossing bond jumps act on its local algebra. Each conjugation
preserves that algebra. Equation~\eqref{eq:auxiliary-exact-channel}
therefore gives
\begin{equation}
 \left\|[(e^{s\mathcal A_g}-\mathcal E_g)(\rho)]_R\right\|_1
 \leq(2M+1)e^{-2s}.
 \label{eq:spin-local-aux-bound}
\end{equation}

After auxiliary averaging, the joint state has the form
\begin{equation}
 \mathcal E_g(\rho)=2^{-N}\bigl[r+P_gX\bigr],
 \label{eq:joint-auxiliary-average}
\end{equation}
where $r$ is an even physical density matrix and $X$ is an even
physical operator. As in Sec.~\ref{app:spins}, physical operators
are embedded in the joint Majorana algebra, and $2^{-N}$ supplies
the normalized auxiliary identity.
The lift identity gives $\mathcal B_p(r)=\mathcal L_p(r)$.
The second term remains $P_g$ times a physical operator and
vanishes on every proper interval. Hence
\begin{equation}
 [e^{s\mathcal B_p}\mathcal E_g(\rho)]_R
 =2^{-M}[e^{s\mathcal L_p}(r)]_R,\qquad
 \sigma_{p,R}=2^{-M}\rho_{\infty,p,R}.
 \label{eq:lifted-physical-evolution}
\end{equation}
This embedding preserves trace norm. The physical mixing bound
\eqref{eq:bogoliubov-mixing} therefore bounds the distance from
$\sigma_{p,R}$ by $\pi M e^{-s/2}$.

Set $\chi_s=e^{s\mathcal B_p}(\rho)$. The triangle inequality now gives
\begin{align}
 \left\|[e^{s\widetilde{\mathcal G}_p}(\rho)]_R-\sigma_{p,R}\right\|_1
 &\leq
 \left\|[(e^{s\mathcal A_g}-\mathcal E_g)(\chi_s)]_R\right\|_1
 \notag\\
 &\quad+
 \left\|[e^{s\mathcal B_p}\mathcal E_g(\rho)]_R-\sigma_{p,R}\right\|_1
 \notag\\
 &\leq(2M+1)e^{-2s}+\pi M e^{-s/2}.
 \label{eq:mixing}
\end{align}
The bound is uniform in $N$ and $p$. Thus every interval relaxes
within error $\varepsilon$ in time $O[\log(M/\varepsilon)]$.

Global parity is conserved, but all parity-even stationary states
have the same proper-interval marginals. In particular,
\begin{equation}
 \omega_\pm=2^{-N}(\id\pm P)\rho_\infty(p),
 \qquad (\omega_+)_R=(\omega_-)_R=\sigma_{p,R},
 \label{eq:spin-stationary-indistinguishability}
\end{equation}
where $\rho_\infty(p)$ is the physical stationary state with
periodic boundary conditions. Its proper-interval marginals
are those found in Sec.~\ref{app:bogoliubov-parent}.
The two states are distinguished by total parity and are locally
indistinguishable.

The auxiliaries only add an extensive entropy,
\begin{equation}
 S(\sigma_{p,R})=S(\rho_{\infty,p,R})+M\log2.
 \label{eq:spin-stationary-entropy}
\end{equation}
This term cancels from the CMI of consecutive complete-cell
regions with $ABC$ a proper interval. Thus
Sec.~\ref{app:bogoliubov-cmi} applies directly: for $0<p<1$, the
lifted stationary states have the same $b^{-4}$ CMI and leading
coefficient as Eq.~\eqref{eq:bogoliubov-general-asymptotic}.

\subsection{Infinite spin chain}

We now use the Jordan--Wigner ordering of Sec.~\ref{app:spins},
with physical qubits at $2j$ and auxiliary qubits at $2j+1$.
On the infinite chain, the jumps become
\begin{align}
 J_j^c&=\sigma^+_{2j}X_{2j+1},&
 J_j^d&=\frac{-Y_{2j}X_{2j+1}+iY_{2j+1}Y_{2j+2}}2,
 \notag\\
 Q_j^z&=Z_{2j+1},&
 Q_j^b&=Y_{2j+1}Z_{2j+2}X_{2j+3},
 \label{eq:spin-steady-jumps}
\end{align}
where $\sigma^+=(X+iY)/2$. These define a translation-invariant,
finite-range spin generator with the same rates.

To include arbitrary spin inputs, average the auxiliary
conjugations meeting an $M$-cell interval $R$ and call this map
$\mathcal E_R$. The onsite averages remove auxiliary occupation
coherences. Commutation with the internal bonds and the two
crossing bonds leaves precisely the Jordan--Wigner images of even
physical Majorana polynomials. Thus
$\mathcal E_R(\rho_R(0))=2^{-M}r_R$ for an even physical
density matrix $r_R$.

Auxiliary averaging commutes with cooling. The physical moment
hierarchy of Sec.~\ref{app:bogoliubov-parent} then gives
\begin{equation}
 \mathcal E_R(\rho_R(s))
 =2^{-M}\Phi_{\alpha(s)}(r_R),
 \label{eq:spin-averaged-physical-channel}
\end{equation}
where $\Phi_{\alpha(s)}$ is the physical channel in
Eq.~\eqref{eq:grassmann-state-identification}.
The distance from $\rho_R(s)$ to its auxiliary average is at
most $(2M+1)e^{-2s}$, while the averaged state is within
$\pi M e^{-s/2}$ of $\sigma_{p,R}$.
The same triangle inequality therefore proves
Eq.~\eqref{eq:mixing} for every initial spin state.
These marginals define the unique infinite-chain spin steady state.
Their spectra obey Eq.~\eqref{eq:spin-stationary-entropy},
so the spin CMI has the same algebraic decay.

\section{Reverse evolution and local recovery for fermions and spins}
\label{app:reverse-and-recovery}

As discussed in Sec.~\ref{sec:implications}, the entire state can follow the trajectory
backward, even though an erased cell cannot be recovered from its
buffer with exponentially small error.

\subsection{Reversing the fermion and spin trajectories}
\label{app:return}

Fix $t_*>0$. We seek loss jumps that take
$N_{t_*}=\id-e^{-t_*K}$ to $N_{t(s)}$, where $t(s)=t_*e^{-s}$.
Define
\begin{equation}
 q_t(x)=\frac{tx}{e^{tx}-1},\qquad
 R_j(s)=\sum_\ell[\sqrt{q_{t(s)}(K)}]_{j\ell}c_\ell,
 \label{eq:return-jumps}
\end{equation}
with $q_t(0)=1$ and $q_0(x)=1$ by continuity. The covariance
under $\sum_j\D{R_j(s)}$ obeys
\begin{equation}
 \partial_sN=-\tfrac12\{q_{t(s)}(K),N\}.
 \label{eq:return-covariance}
\end{equation}
The choice of $q_t$ follows from
\begin{equation}
 q_{t(s)}(K)N_{t(s)}=t(s)K e^{-t(s)K}
 =-\partial_sN_{t(s)}.
 \label{eq:return-identity}
\end{equation}
Since $N_t$ commutes with $q_t(K)$, this identity proves that
$N_{t(s)}$ solves Eq.~\eqref{eq:return-covariance}. Linear loss
preserves Gaussianity, so the full state is exactly
$\rho_{\rm rev}(s)=\rho_{t_*e^{-s}}$.

The jumps are exponentially local. On the spectrum
$0\leq x\leq4$ of $K$, $0<q_t(x)\leq1$, and its positive square
root extends analytically to a common neighborhood for
$0\leq t\leq t_*$. Since $K$ is nearest neighbor, the matrix
elements of $\sqrt{q_t(K)}$ decay exponentially with distance,
with bounds uniform over this time interval. In the spin
construction, replace $R_j$ by $i\gamma_jR_j$, using the auxiliary
Majorana at cell $j$. Each term then has a Jordan--Wigner string
only between $j$ and $\ell$, with an exponentially decaying
coefficient. The embedding identity of Sec.~\ref{app:spins} therefore gives the same exact return for the
spin state, with exponentially local spin jumps.

Every earlier state with $0<t'<t_*$ is reached at
$s=\log(t_*/t')$. As $s\to\infty$, the jumps approach onsite loss
and the state approaches the initial product state. These statements
apply to the infinite chain and to a finite open chain when the
same full finite-chain matrix $K$ defines both evolutions. 

\subsection{Limits on local recovery in the spin model}
\label{app:erasure-recovery}

Let $\rho$ be the exact spin state on $b+2$ consecutive cells.
Take $A$ and $C$ to be the two end cells and $B$ the $b$ cells
between them. Erase $A$, and let an arbitrary channel
$\mathcal R_{B\to AB}$ act on the buffer to reconstruct it. Write
\begin{equation}
 \tau=(\mathcal R_{B\to AB}\otimes\mathrm{id}_C)(\rho_{BC}),
 \qquad \delta=\tfrac12\|\tau-\rho\|_1.
\end{equation}
Data processing gives $I_\tau(AB:C)\leq I_\rho(B:C)$.
We also have $\tau_C=\rho_C$, since the channel does not act on $C$.
Together these give
\begin{equation}
 I_\rho(A:C|B) \leq I_\rho(AB:C)-I_\tau(AB:C) =S_\tau(C|AB)-S_\rho(C|AB),
\end{equation}
where $S(C|AB)=S(ABC)-S(AB)$. The continuity bound for
conditional entropy~\cite{Winter2016Continuity} gives
\begin{equation}
 I_\rho(A:C|B)\leq
 2\delta\log d_C+(1+\delta)h\!\left(\frac{\delta}{1+\delta}\right),
 \label{eq:erasure-error-cmi}
\end{equation}
with $d_C=4$ for one complete spin cell. The dimension in this
bound is independent of the buffer length.

At fixed $t>0$, Sec.~\ref{app:finite-time} gives a
positive lower bound proportional to $b^{-4}$. If a family of
recovery channels had $\delta\leq C e^{-b/\xi}$, with $C$ and
$\xi$ independent of $b$, the right side of
Eq.~\eqref{eq:erasure-error-cmi} would instead be
$O(b e^{-b/\xi})$, a contradiction. Thus increasing the buffer
cannot give exponentially accurate recovery of the erased cell.
The return above acts throughout the chain, so it is compatible
with this obstruction to a repair supported on the buffer.

\subsection{Accuracy of phase equivalence}
\label{app:superpolynomial-accuracy}

We write \(\epsilon_L=L^{-\infty}\) when
\(L^n\epsilon_L\to0\) for every fixed \(n>0\).
This accuracy retains polynomially weak correlations at macroscopic
separation.  To illustrate the distinction, consider a chain of
\(L\) qubits in the state
\begin{equation}
 \begin{aligned}
 |\psi_L\rangle&=\sqrt{1-\delta_L^2}|0^L\rangle
                  +\delta_L|W_L\rangle,
 &\delta_L&=L^{-\kappa},\\
 |W_L\rangle&=\frac1{\sqrt L}\sum_{j=1}^L|j\rangle,
 &\kappa&>0,
 \end{aligned}
 \label{eq:small-w-state}
\end{equation}
where \(|j\rangle\) has one excitation at site \(j\).
Its density matrix is at trace-norm distance \(2\delta_L\) from
\(|0^L\rangle\langle0^L|\), so the identity channel in both
directions would make these families equivalent at merely polynomially vanishing
accuracy.  Yet for any distinct sites,
\begin{equation}
 \begin{aligned}
 C_X^\psi(i,j)
 &=\langle X_iX_j\rangle_\psi
  -\langle X_i\rangle_\psi\langle X_j\rangle_\psi\\
 &=-\frac{2\delta_L^2(1-2\delta_L^2)}{L}.
 \end{aligned}
 \label{eq:small-w-connected-correlation}
\end{equation}
This polynomial correlation is independent of their separation.

A state \(\tau_L\) prepared from a product input by a local circuit
with light-cone radius \(o(L)\) has zero connected correlation between
sites a distance \(\Theta(L)\) apart: their backward light cones are
disjoint.  Connected correlations of norm-one observables change by
at most three times the trace-norm error.  To see this,
write
\(C_\sigma(A,B)=\langle AB\rangle_\sigma
-\langle A\rangle_\sigma\langle B\rangle_\sigma\).
For \(\|A\|_\infty,\|B\|_\infty\leq1\), adding and subtracting
\(\langle A\rangle_\tau\langle B\rangle_\rho\) gives
\begin{equation}
 \begin{aligned}
 |C_\rho(A,B)-C_\tau(A,B)|
 &\leq
 |\langle AB\rangle_\rho-\langle AB\rangle_\tau|\\
 &\quad+
 |\langle A\rangle_\rho-\langle A\rangle_\tau|\,
 |\langle B\rangle_\rho|\\
 &\quad+
 |\langle A\rangle_\tau|\,
 |\langle B\rangle_\rho-\langle B\rangle_\tau|\\
 &\leq 3\|\rho-\tau\|_1.
 \end{aligned}
 \label{eq:connected-correlation-continuity}
\end{equation}
Here the last line follows from
\(
 |\operatorname{tr}[(\rho-\tau)O]|
 \leq\|\rho-\tau\|_1\|O\|_\infty
\)
and the fact that all one-point expectations have magnitude at
most one. Consequently,
\begin{equation}
 \||\psi_L\rangle\langle\psi_L|-\tau_L\|_1
 \geq \frac{2\delta_L^2(1-2\delta_L^2)}{3L}
 =\Omega(L^{-(2\kappa+1)}).
 \label{eq:small-w-circuit-lower-bound}
\end{equation}
Thus no such channel prepares \(|\psi_L\rangle\) with
superpolynomial accuracy.

The distinction also has a Hamiltonian interpretation.  The product
state has a uniformly gapped local parent Hamiltonian, whereas
Eq.~\eqref{eq:small-w-connected-correlation} is incompatible with
exponential clustering for a unique ground state of a uniformly
gapped, bounded-strength, finite-range Hamiltonian
~\cite{HastingsKoma2006}.  The state \(|\psi_L\rangle\) does belong to
the symmetric ground space of the ferromagnetic chain
\(H_{\rm F}=\frac12\sum_{i=1}^{L-1}
(\mathbb I-\operatorname{SWAP}_{i,i+1})\),
whose one-magnon gap closes as \(1-\cos(\pi/L)\).
Conversely, uniformly gapped paths admit local-circuit approximations
to quasi-local continuation with polylogarithmic range and
superpolynomially small error~\cite{Coser2019}.

\section{CMI decay under approximate local reversibility}
\label{app:approx-local-reversibility}

We follow the proof of Yi et al.~\cite{Yi2026}, allowing
the local reverse maps to have a small error and the light
cone to grow with system size. The idea is the same as in
their exact argument. Suppose $\sigma=\mathcal E(\rho)$
and we want to recover an erased region $C$ of $\sigma$
using only the buffer $B$, without acting on $A$.
We first reverse the evolution, recover the missing part
of $\rho$ with a Petz map, and evolve forward again:
\begin{equation}
 \mathcal Q=\mathcal E\circ\mathcal P\circ\mathcal R.
 \label{eq:lr-sandwich-picture}
\end{equation}
This formula describes the order of the three operations.
The actual recovery uses only the parts of the forward
and reverse evolution that can be performed from $B$.
Reversing the evolution enlarges the erased region, so
the Petz map must recover this larger region. The gates
near $A$ cancel by local reversibility, leaving a map
that does not act on $A$. This is the construction behind
Eqs.~(VII.14)--(VII.18) of Ref.~\cite{Yi2026}.

We carry this single Petz map through the circuit one
layer at a time. The enlargement of the regions then
adds up to the total light-cone radius, and the errors
of the reverse maps add in trace distance. In particular,
we use the relation between CMI and recovery only at
the beginning and the end of the proof.
When the total light cone is sublinear and the total
error is superpolynomially small, the result preserves
superpolynomial CMI across buffers whose width is
proportional to the system size.

Consider a lattice with $O(L^D)$ sites and a fixed local
Hilbert-space dimension. Write the evolution as
\begin{equation}
 \rho_t=\mathcal E_t(\rho_{t-1}),\qquad
 \rho_0=\rho,\qquad t=1,\ldots,T.
 \label{eq:lr-trajectory}
\end{equation}
Each layer consists of channels on disjoint blocks,
with identity channels on the remaining sites. If
the blocks in layer $t$ have diameter at most $r_t$,
the total light-cone radius is bounded by
\begin{equation}
 r=\sum_{t=1}^T r_t.
 \label{eq:lr-total-cone}
\end{equation}
For a union $Y$ of blocks in this layer, denote the
product of its gates by $\mathcal E_t^Y$. We assume
there is a channel $\mathcal R_t^Y$ acting in $Y$ such that
\begin{equation}
 d\!\left(\mathcal R_t^Y\mathcal E_t^Y(\rho_{t-1}),
             \rho_{t-1}\right)\leq\delta_t,
 \qquad d(\alpha,\beta)=\tfrac12\|\alpha-\beta\|_1.
 \label{eq:lr-regional-reversal}
\end{equation}
Here and below identities outside the indicated regions
are implicit. The error in this assumption is an error
on the full state: the reverse must also restore
correlations with the complement of $Y$. We only need
the assumption for the regions that occur in the proof.
It is the approximate form of Definition 1 in
Ref.~\cite{Yi2026}.

This condition follows, for example, from reversibility
of each gate on its own block. If these errors on
$\rho_{t-1}$ are $\epsilon_{t,j}$, the channels on
different blocks commute, and telescoping their product
gives $\delta_t\leq\sum_j\epsilon_{t,j}$ by contractivity.
A reverse map may act on a larger region if the gates
can be grouped into disjoint blocks containing these
supports; the larger diameters must then be included
in $r$. A reverse of the whole circuit alone does not
give Eq.~\eqref{eq:lr-regional-reversal}.

Let $A,B,C$ partition the lattice, with $B$ separating
$A$ and $C$, and suppose $b=\operatorname{dist}(A,C)>2r$.
Starting with $A_T=A$ and $C_T=C$, work backwards through
the circuit. At each step, enlarge $A_t$ and $C_t$ to
include every block of layer $t$ that meets them, and
call the resulting regions $A_{t-1}$ and $C_{t-1}$.
Their complement is $B_{t-1}$. The two endpoints remain
disjoint, each grows by at most $r$ in total, and hence
\begin{equation}
 \operatorname{dist}(A_0,C_0)\geq b-2r.
 \label{eq:lr-cone-geometry}
\end{equation}
The initial recovery map acts from $B_0$ to $B_0C_0$.
Set $I_0=I(A_0:C_0|B_0)_\rho$. The recovery
theorem~\cite{FawziRenner2015,Junge_2018}
gives a Petz map $\mathcal P$, averaged over modular
rotations, for which
\begin{equation}
 d\!\left(\rho,
   (\mathrm{id}_{A_0}\otimes\mathcal P)(\rho_{A_0B_0})\right)
 \leq\sqrt{1-e^{-I_0}}\leq\sqrt{I_0}.
 \label{eq:lr-initial-petz}
\end{equation}
We use natural logarithms. This bound follows from the
fidelity estimate $F\geq e^{-I_0/2}$ and the
fidelity--trace-distance inequality. This estimate is
for the averaged recovery map. On the kernel of the input marginal,
choose a trace-preserving extension.

Now suppose $\mathcal Q_{t-1}$ recovers $C_{t-1}$ from
$B_{t-1}$ in $\rho_{t-1}$ with error at most $u_{t-1}$.
To see how to move this recovery through the next
layer, write its three input regions as
\begin{equation}
 X=A_{t-1}=A_t x,\qquad Y=B_{t-1},\qquad
 Z=C_{t-1}=zC_t,\qquad B_t=xYz.
 \label{eq:lr-one-layer-regions}
\end{equation}
The regions $x$ and $z$ are the parts of the
buffer absorbed when following this layer backwards.
Because $X,Y,Z$ are unions of complete blocks, the
layer factors as
$\mathcal E_t=\mathcal E_t^X\otimes
\mathcal E_t^Y\otimes\mathcal E_t^Z$.
Define
\begin{equation}
 \mathcal Q_t
 =(\mathcal E_t^Y\otimes\mathcal E_t^Z)
       \circ\mathcal Q_{t-1}
       \circ\mathcal R_t^Y\circ\Tr_z,
 \label{eq:lr-localized-sandwich}
\end{equation}
with the identity on $x$ understood. This is the
Petz sandwich after one layer. It discards $z$, undoes
the evolution on $Y$, recovers $Z$, and evolves $YZ$
forward again. Its input is $B_t=xYz$, its output is
$B_tC_t=xYZ$, and it never acts on $A_t$.

In the exact proof, the forward and reverse gates on
$X$ can be removed because
\begin{equation}
 \mathcal E_t^X\mathcal R_t^X((\rho_t)_{XY}) =\mathcal E_t^X\mathcal E_t^Y \bigl(\mathcal R_t^X\mathcal E_t^X ((\rho_{t-1})_{XY})\bigr) =(\rho_t)_{XY}.
 \label{eq:lr-state-cancellation}
\end{equation}
The equality only uses reversibility on the state at
this step. It does not require the channels to be
inverses on every state. This is the cancellation used
in Eq.~(VII.18) of Ref.~\cite{Yi2026}.
Equation~\eqref{eq:lr-localized-sandwich} already omits
these gates, and its error can be checked directly.

Put $\eta=\rho_{t-1}$ and $\tau=\rho_t$. Tracing out
$C_t$ and then $z$ removes all of $Z$, so
\begin{equation}
 \Tr_z\tau_{A_tB_t}
 =\tau_{XY}
 =(\mathcal E_t^X\otimes\mathcal E_t^Y)(\eta_{XY}).
\end{equation}
The operations on $X$ commute with the recovery
operations on $YZ$. Substituting this identity into
Eq.~\eqref{eq:lr-localized-sandwich} gives
\begin{equation}
 (\mathrm{id}_{A_t}\otimes\mathcal Q_t)(\tau_{A_tB_t}) =\mathcal E_t \left[(\mathrm{id}_X\otimes\mathcal Q_{t-1}) \left((\mathrm{id}_X\otimes \mathcal R_t^Y\mathcal E_t^Y)(\eta_{XY})\right)\right].
 \label{eq:lr-transport-identity}
\end{equation}
By Eq.~\eqref{eq:lr-regional-reversal}, replacing the
innermost state by $\eta_{XY}$ changes the result by
at most $\delta_t$. Applying $\mathcal Q_{t-1}$ to
$\eta_{XY}$ recovers $\eta$ within $u_{t-1}$, and the
last forward layer cannot increase this error. Thus
\begin{equation}
 u_t\leq u_{t-1}+\delta_t.
 \label{eq:lr-error-recurrence}
\end{equation}
This is the change needed for approximate reversibility:
the exact equality is replaced by an additive error.
The reverse is applied to the actual intermediate
state, before recovery, so no assumption about its
action on other states is used.

Starting with $\mathcal Q_0=\mathcal P$, we obtain a
channel $\mathcal Q_T:B\to BC$ with error at most
$\sqrt{I_0}+\delta$, where $\delta=\sum_t\delta_t$.
If the desired final state $\sigma$ also has a
preparation error $d(\sigma,\rho_T)\leq\epsilon$,
the same recovery has error at most
\begin{equation}
 u=\min\{1,\sqrt{I_0}+\delta+2\epsilon\}.
 \label{eq:lr-error-budget}
\end{equation}
The two endpoint errors come from replacing the input
marginal and the target state. Both are bounded by
$\epsilon$ using contractivity and the triangle inequality.

It remains to turn this recovery estimate back into a
CMI bound. Let
$\widehat\sigma=(\mathrm{id}_A\otimes\mathcal Q_T)(\sigma_{AB})$.
Data processing gives
$I(A:BC)_{\widehat\sigma}\leq I(A:B)_\sigma$,
and $\widehat\sigma_A=\sigma_A$. Therefore
\begin{equation}
 I(A:C|B)_\sigma \leq I(A:BC)_\sigma-I(A:BC)_{\widehat\sigma} =S(A|BC)_{\widehat\sigma}-S(A|BC)_\sigma.
 \label{eq:lr-recovery-converse}
\end{equation}
The continuity bound for conditional
entropy~\cite{Winter2016Continuity} then gives
\begin{equation}
 I(A:C|B)_\sigma
 \leq 2u\log d_A+(1+u)h\!\left(\frac{u}{1+u}\right),
 \label{eq:lr-cmi-bound}
\end{equation}
where $d_A=\dim\mathcal H_A$ and
$h(v)=-v\log v-(1-v)\log(1-v)$.
This argument does not require $\widehat\sigma$ to be
an exact quantum Markov state.
For $d_A\geq2$, the right side is at most
$7\log d_A\sqrt u$: indeed,
$(1+u)h(u/(1+u))\leq2u-u\log u\leq3\sqrt u$
for $0\leq u\leq1$.
Consequently,
\begin{equation}
 I(A:C|B)_\sigma
 \leq7\log d_A\left[I_0^{1/4}+
                         \sqrt{\delta+2\epsilon}\right].
 \label{eq:lr-quarter-power}
\end{equation}
This has the same fourth root of the initial CMI as
Eq.~(VII.19) of Ref.~\cite{Yi2026}, with an additional
term from imperfect reversal and preparation.
Thus the Petz step of their proof is unchanged; the
regions grow by the sum of the layer radii, and the
reverse-map errors add before the final conversion
back to CMI.

To state the decay consequence, suppose the initial
state satisfies
\begin{equation}
 I(X:Z|Y)_\rho
 \leq P(|X|,|Y|)\,f(\operatorname{dist}(X,Z)),
 \qquad f(s)=s^{-\omega(1)},
 \label{eq:lr-initial-decay}
\end{equation}
uniformly on the enlarged partitions used above.
Here $P$ is a polynomial, and $f$ is nonincreasing
and decays faster than every inverse power.
The regions may also be enlarged to regular
neighborhoods: moving sites from the buffer to either
endpoint can only increase CMI, by the chain rule
and strong subadditivity. Equations
\eqref{eq:lr-cone-geometry} and \eqref{eq:lr-quarter-power}
give
\begin{equation}
 I(A:C|B)_\sigma\leq7\log d_A\bigl[ P(|A_0|,|B_0|)^{1/4}f(b-2r)^{1/4} +\sqrt{\delta+2\epsilon}\bigr].
 \label{eq:lr-shifted-decay}
\end{equation}
If $r=o(L)$, $\delta+\epsilon=L^{-\omega(1)}$,
and $b\geq cL$ for fixed $c>0$, then $b-2r\geq cL/2$
for large $L$. All volume factors and $\log d_A$ grow
at most polynomially in $L$. Taking a fixed positive
power and multiplying by such factors preserves
superpolynomial decay, so
\begin{equation}
 I(A:C|B)_\sigma=L^{-\omega(1)}.
 \label{eq:lr-macroscopic-invariance}
\end{equation}
The same conclusion holds at intermediate times.
If the assumptions also hold along an evolution in
the opposite direction, the implication holds both ways.
This gives the claimed invariance.

The assumptions concern the total radius and total
error. A polynomial number of gates, each with a
uniformly superpolynomial reversal error, is sufficient.
When $\delta$ is bounded directly, no additional
restriction on the number of layers is needed.
More generally, Eq.~\eqref{eq:lr-shifted-decay} preserves
superpolynomial decay in the buffer width whenever
$b-2r\geq\kappa b$ for a fixed $\kappa>0$ and the
volume factors are polynomial in $b$.
If also $b\geq L^\beta$ for fixed $\beta>0$, this gives
superpolynomial decay in $L$.
The restriction to distances beyond the light cone
is necessary. An exactly reversible circuit can
create a Bell pair and move its halves to a separation
of order $\sqrt L$ in depth $O(\sqrt L)$ using
nearest-neighbor unitaries. Its CMI, conditioned on
the remaining sites, is $2\log2$, even though the
initial state was a product. A sublinear light cone
therefore allows correlations at distances of its own
size.

Finally, the proof applies when $ABC$ is a proper
interval of the chain, as in this paper. Take
consecutive intervals $A,B,C$ and contiguous gate
blocks, with enough exterior space that the enlarged
interval does not wrap around the ring. At each
backward step include every block meeting the
interval. The extra exterior sites belong to the
left or right endpoint; they never enter the middle
region $Y$. In the notation of
Eq.~\eqref{eq:lr-one-layer-regions}, write
$X=A_txe_X$ and $Z=zC_te_Z$, where $e_X,e_Z$
are these exterior sites. Replace the endpoint
channels by $\Tr_{e_X}\circ\mathcal E_t^X$ and
$\Tr_{e_Z}\circ\mathcal E_t^Z$.
The left trace commutes with recovery on $YZ$,
and the right trace is performed after recovery.
Contractivity therefore gives the same recurrence
\eqref{eq:lr-error-recurrence}. Each endpoint still
grows by at most $r$, and the buffer shrinks by at
most $2r$.
For a product input, $I_0=0$ on every enlarged
partition. The positive algebraic CMI lower bounds
in Secs.~\ref{app:finite-time} and
\ref{app:bogoliubov-cmi}, with a proper buffer of
width $\Theta(L)$, thus rule out preparation from
a product state by circuits with these local
reverses, a sublinear total light cone, and a
superpolynomially small total error.

\section{Ordinary and nonlinear correlations in the fermion and spin models}
\label{app:correlations}

\subsection{Density correlations and the canonical purification}
\label{app:nonlinear}

We compare the ordinary connected density correlation with
\begin{equation}
 W_t(r)=\Tr(\sqrt{\rho_t}\,n_0\sqrt{\rho_t}\,n_r)
       -\langle n_0\rangle\langle n_r\rangle,
 \qquad n_j=c_j^\dagger c_j.
 \label{eq:wightman}
\end{equation}
The square-root trace is first evaluated for the state on a finite
chain. We take the bulk thermodynamic limit at fixed $r$, and then
let $r$ grow. In particular, the square root is taken before any
restriction to the two measured sites.

The ordinary correlation is short ranged. With the main-text
symbol $\kappa(k)=2+2\cos k$, the Fourier coefficients of
$e^{-t\kappa(k)}$ give, for $r\ne0$,
\begin{equation}
 N_{0r}=-(-1)^r e^{-2t}\mathsf I_r(2t),\qquad
 \langle n_0n_r\rangle_c=-|N_{0r}|^2,
 \label{eq:ordinary-density-correlation}
\end{equation}
where $\mathsf I_r$ is the modified Bessel function and the second
identity is Wick's rule. At fixed $t$, its series gives
$\mathsf I_r(2t)\sim t^r/r!$, so this correlation decays faster
than exponentially.

The square-root trace has the Gaussian formula:
\begin{equation}
 W(i,j)=\left|[\sqrt{N(\id-N)}]_{ij}\right|^2.
 \label{eq:w-gaussian}
\end{equation}
To derive it, diagonalize $N$ into independent modes $f_m$
with occupations $\lambda_m$, and put
$s_m=\sqrt{\lambda_m(1-\lambda_m)}$. Taking the trace in their occupation basis gives
\begin{equation}
 \Tr(\sqrt\rho\,f_m^\dagger f_n\sqrt\rho\,f_p^\dagger f_q) =\delta_{mn}\delta_{pq}\lambda_m\lambda_p +\delta_{mq}\delta_{np}s_ms_n.
 \label{eq:w-mode-trace}
\end{equation}
The first term comes from unchanged occupations. The second
exchanges a particle between two modes and multiplies the square
roots of their occupation probabilities. When all four indices
coincide, the sum is $\lambda_m^2+s_m^2=\lambda_m$, as required
by $n_m^2=n_m$. Expanding the site densities in these modes and
subtracting their product expectation proves
Eq.~\eqref{eq:w-gaussian}, including pure modes by continuity.

The Fourier symbol in Eq.~\eqref{eq:w-gaussian} is
\begin{equation}
 w_t(k)=e^{-t\kappa(k)/2}\sqrt{1-e^{-t\kappa(k)}}.
\end{equation}
Unlike the ordinary covariance, it has a cusp:
$w_t(\pi+q)=\sqrt t\,|q|+O_t(|q|^3)$. It is smooth away
from $k=\pi$. Two integrations by parts show that
the derivative jump $2\sqrt t$ gives
\begin{equation}
 [\sqrt{N_t(\id-N_t)}]_{0r}
 =\frac{(-1)^{r+1}\sqrt t}{\pi r^2}+O_t(r^{-4}).
\end{equation}
The next term comes from the cubic cusp. Squaring therefore yields
\begin{equation}
 W_t(r)=\frac{t}{\pi^2r^4}+O_t(r^{-6}).
 \label{eq:nonlinear-tail}
\end{equation}

This calculation also applies to the local spin densities
$n_j=(1-Z_{2j})/2$. In the notation of Sec.~\ref{app:spins}, the normalized embedded state is
$\widehat\rho=2^{-M}r(\eta,\xi)$, and functional calculus gives
$\sqrt{\widehat\rho}=2^{-M/2}\sqrt r$. The trace over the
auxiliaries contributes $2^M$, canceling the two normalization
factors. Thus the square-root trace of even physical observables
is preserved, just as their ordinary expectations are.

The normalized vector $|\sqrt\rho\rangle\rangle$
obtained by vectorizing $\sqrt\rho$ is the canonical purification.
For site densities,
\begin{equation}
 \langle\langle\sqrt\rho|
 n_i\otimes n_j^{\mathsf T}|\sqrt\rho\rangle\rangle
 =\Tr(\sqrt\rho\,n_i\sqrt\rho\,n_j).
\end{equation}
Density is even and diagonal, so this identity has no fermionic
or transpose sign. Equation~\eqref{eq:nonlinear-tail} is therefore
an algebraic connected correlation between the two copies.

\subsection{Connection to fidelity correlators}
\label{app:fidelity-correlations}

To compare Eq.~\eqref{eq:wightman} with the nonlinear order
parameters of Refs.~\cite{LessaEtAl2025,LiuEtAl2025,LiFirankoHsieh2026},
let $\|O\|\leq1$ and define
\begin{equation}
 F_O=\|\sqrt\rho O\sqrt\rho\|_1,\qquad C_W(O)=\Tr(\sqrt\rho O\sqrt\rho O^\dagger).
\end{equation}
Here $F_O=F(\rho,O\rho O^\dagger)$ uses the unsquared fidelity
$F(\rho,\sigma)=\Tr\sqrt{\sqrt\rho\sigma\sqrt\rho}$;
this formula also defines it when $O\rho O^\dagger$ has trace
less than one. The bounds are
\begin{equation}
 C_W(O)\leq F_O\leq\sqrt{C_W(O)}.
 \label{eq:fidelity-wightman-bounds}
\end{equation}
The first follows from trace-norm duality and $\|O\|\leq1$.
For the second, write
$\sqrt\rho O\sqrt\rho=\rho^{1/4}
(\rho^{1/4}O\rho^{1/4})\rho^{1/4}$ and apply the Schatten
H\"older inequality with exponents $4,2,4$. The outer two norms
are one and the middle norm is $\sqrt{C_W(O)}$.

For $O=O_iO_j^\dagger$, these bounds relate the nonzero
large-distance limits used as symmetry-breaking diagnostics in
Ref.~\cite{LiuEtAl2025}. Our $W_t(r)$ instead places one density
on each copy and subtracts the product expectation. Its algebraic
tail alone therefore neither fixes a fidelity-correlator exponent
nor establishes symmetry breaking.

It does determine an overlap response. Apply two density
rotations,
$U(\theta,\varphi)=e^{i\theta n_0+i\varphi n_r}$, and define
\begin{equation}
 \mathcal A_t(\theta,\varphi;r)
 =\Tr[\sqrt{\rho_t}\,U\sqrt{\rho_t}\,U^\dagger].
 \label{eq:overlap-response-definition}
\end{equation}
For unitary $U$ this is the square-root overlap
$\Tr[\sqrt{\rho_t}\sqrt{U\rho_tU^\dagger}]$, so
Eq.~\eqref{eq:fidelity-wightman-bounds} gives
$\mathcal A_t\leq F(\rho_t,U\rho_tU^\dagger)
\leq\sqrt{\mathcal A_t}$. Since the two densities commute,
differentiating the two factors of $U$ gives
\begin{equation}
 \left.\partial_\theta\partial_\varphi\mathcal A_t\right|_{0} =2\bigl[\Tr(\sqrt{\rho_t}n_0\sqrt{\rho_t}n_r) -\langle n_0n_r\rangle\bigr] =2\bigl[W_t(r)-\langle n_0n_r\rangle_c\bigr] =\frac{2t}{\pi^2r^4}+O_t(r^{-6}).
 \label{eq:overlap-response}
\end{equation}
Thus the response couples distant density rotations algebraically,
although their ordinary connected correlation is short ranged.
The identity is unchanged for the local densities in the spin model.

\subsection{Power-law CMI with a short-range canonical purification}
\label{app:short-range-purification}

We give a local fermionic evolution whose CMI has a power-law
lower bound while its canonical purification remains short ranged.
Starting from the vacuum, take
$\dot\rho_t=\sum_j[\D{F_j}(\rho_t)+\D{G_j}(\rho_t)]$, with
\begin{equation}
 F_j=\frac{\sqrt3}{8}
 (c_j^\dagger-2c_{j+1}^\dagger+c_{j+2}^\dagger),
 \qquad
 G_j=\frac18(3c_j+6c_{j+1}-c_{j+2}).
 \label{eq:quartic-jumps}
\end{equation}
Write $\kappa(k)=|1-e^{ik}|^4=(2-2\cos k)^2$.
The gain and loss rates are $3\kappa(k)/64$ and
$|3+6e^{ik}-e^{2ik}|^2/64=1-3\kappa(k)/64$.
Thus $\partial_t n=3\kappa/64-n$, and
\begin{equation}
 n(k,t)=q_t\kappa(k),\qquad
 q_t=\frac3{64}(1-e^{-t}),\qquad N_t=q_tK,
 \label{eq:quartic-covariance}
\end{equation}
where $K_{ij}=6\delta_{ij}-4\delta_{|i-j|,1}
+\delta_{|i-j|,2}$. In particular, the physical mutual information
vanishes between regions separated by more than two sites.

The canonical-purification kernel has symbol
\begin{equation}
 w_t(k)=\sqrt{n(k,t)[1-n(k,t)]}
 =\sqrt{q_t}(2-2\cos k)\sqrt{1-q_t\kappa(k)}.
 \label{eq:quartic-purification}
\end{equation}
Since $q_t\kappa(k)\leq3/4$ on the real axis, this function is
analytic in a strip around it. Its Fourier coefficients therefore
decay exponentially. The purified Gaussian covariance consists
of $N_t$, $\id-N_t$, and $\sqrt{N_t(\id-N_t)}$, up to a fixed
particle-hole convention on the auxiliary copy, so it is
exponentially local. This also gives exponential mutual-information
decay for separated finite blocks: deleting the cross block changes
the covariance by an exponentially small norm, and
$|h(x)-h(y)|\leq C|x-y|^{1/2}$ bounds the entropy change.
Equation~\eqref{eq:w-gaussian} gives exponential decay of $W_t(r)$
as well.

To bound the CMI, take $a=c=1$, $b\geq1$, and $L=b+2$.
Let $K_L$ be the principal submatrix on $ABC$.
Using Eq.~\eqref{eq:creation-cmi-positive-integral}, keeping its
first nonnegative term, and rescaling the integration variable gives
\begin{equation}
 I(A:C|B;t)\geq q_t\int_0^\infty du\,J(K_L+u\id),
 \label{eq:quartic-cmi-integral}
\end{equation}
where $J$ is defined in Eq.~\eqref{eq:creation-logdet-cmi}.
Set $R(u)=(K_L+u\id)^{-1}$. Taking the Schur complement of
the buffer gives
\begin{equation}
 J(K_L+u\id)
 =-\log\left(1-\frac{|R(u)_{1L}|^2}{R(u)_{11}R(u)_{LL}}\right)
 \geq\frac{|R(u)_{1L}|^2}{R(u)_{11}R(u)_{LL}}.
 \label{eq:quartic-inverse-cmi}
\end{equation}
The first column of $R(0)$ is
\begin{equation}
 R(0)_{j1}=\frac{j(L+1-j)(L+2-j)}{(L+2)(L+3)},
 \qquad 1\leq j\leq L,
 \label{eq:quartic-inverse-column}
\end{equation}
as one checks by multiplying by $K_L$. For $L\geq3$, this gives
$R(0)_{1L}\geq1/(2L)$, $R(0)_{11}<1$, and
$\|R(0)e_1\|_2^2\leq L^3$. This implies
\begin{equation}
 |R(u)_{1L}-R(0)_{1L}|
 \leq u\|R(0)e_1\|_2\|R(0)e_L\|_2
 \leq uL^3.
 \label{eq:quartic-resolvent}
\end{equation}
For $0\leq u\leq1/(4L^4)$, it follows that
$R(u)_{1L}\geq1/(4L)$. Also $R(u)\leq R(0)$, so its endpoint
diagonal entries are at most one.
Equation~\eqref{eq:quartic-inverse-cmi} is therefore at least
$1/(16L^2)$ throughout this interval. Integration yields
\begin{equation}
 I(A:C|B;t)\geq\frac{q_t}{64(b+2)^6}.
 \label{eq:quartic-cmi-lower}
\end{equation}

The bound also holds with a two-sided buffer. Take $A$ to contain
two sites, put $b$ buffer sites $B_-$ and $B_+$ on either side,
and retain one exterior site $C_+$ just beyond $B_+$.
The range-two covariance makes $B_-$ independent of $B_+C_+$.
The chain rule and strong subadditivity then give
\begin{equation}
 I(A:C_+|B_-B_+)
 =I(A:C_+|B_+)+I(B_-:C_+|AB_+)
 \geq I(A:C_+|B_+).
 \label{eq:quartic-two-sided}
\end{equation}
Discarding the left site of $A$ reduces the last term to the
geometry of Eq.~\eqref{eq:quartic-cmi-lower}. Enlarging $C_+$ to
the full exterior can only increase the CMI. Thus the Markov
length is infinite for every $t>0$, while the canonical
purification remains short ranged.

\section{Vanishing CMI at large scales in the fermion and spin models}
\label{app:infrared}

At fixed $t>0$, we show that the interval entropy has the form
\begin{equation}
 S_L=Ls(t)+\gamma(t)+o(1),\qquad
 s(t)=\int_{-\pi}^{\pi}\frac{dk}{2\pi}h[n(k)],
 \label{eq:creation-entropy-edge-limit}
\end{equation}
with finite $\gamma(t)$. This proves both infrared limits in the
main text, since the volume and constant terms cancel from the CMI.
We control the limit without assuming an entropy power series near
the zero of $n(k)$.

Keep $\kappa(k)=2+2\cos k$, and write
$n(k)=1-e^{-t\kappa(k)}$ and $q(k)=1-n(k)$.
For a symbol $w$, let $T_L(w)$ be its $L\times L$ Toeplitz
matrix. The interval covariance is exactly $T_L(n)$.
For smooth positive $w$, define
\begin{equation}
 \ell_j(w)=\int_{-\pi}^{\pi}\frac{dk}{2\pi}e^{-ijk}\log w(k),\qquad E_L(w)=\log\det T_L(w)-L\ell_0(w),\qquad E(w)=\sum_{j\geq1}j|\ell_j(w)|^2.
 \label{eq:creation-szego-boundary}
\end{equation}
The strong Szeg\H{o} theorem, together with monotonicity of the
normalized Toeplitz determinant, gives
$0\leq E_L(w)\uparrow E(w)$; see Theorems~1.1 and 2.4 of
Ref.~\cite{Simon2004StrongSzego}. We use this only for the
strictly positive symbols $n+u$ and $q+u$, with $u>0$.
The entropy identity in Eq.~\eqref{eq:creation-entropy-integral}
then gives exactly
\begin{equation}
 S_L-Ls(t)=\int_0^\infty du\,[E_L(n+u)+E_L(q+u)].
 \label{eq:creation-entropy-boundary-integral}
\end{equation}
We next show that the limiting integrand is integrable, which
justifies interchanging the entropy limit and integral.

Only $u=0$ is potentially singular, because $n$ has a quadratic
zero. For $0<u\leq1$, put $v=t^{-1}\log(1+u)$ and factor
\begin{equation}
 n(k)+u=[\kappa(k)+v]g_u(k),\qquad g_u(k)=(1+u)\frac{1-e^{-t[\kappa(k)+v]}}{\kappa(k)+v}.
 \label{eq:creation-zero-factorization}
\end{equation}
The apparent singularity of $g_u$ is removable. At fixed $t$,
$\log g_u$ and its first derivative are uniformly bounded for
$0<u\leq1$. Parseval's identity therefore bounds $E(g_u)$ by
a constant depending only on $t$.

For the singular factor, set $r=e^{-\eta}$
with $2\cosh\eta=2+v$. Then
\begin{equation}
 \kappa(k)+v=r^{-1}(1+re^{ik})(1+re^{-ik}),\qquad \ell_j(\kappa+v)=-\frac{(-r)^{|j|}}{|j|}\quad(j\ne0),\qquad E(\kappa+v)=-\log(1-r^2).
\end{equation}
Since $\eta\sim\sqrt{u/t}$ as $u\downarrow0$, this divergence
is logarithmic and hence integrable. The bound
$E(n+u)\leq2E(\kappa+v)+2E(g_u)$ controls the full symbol.
The other symbol stays away from zero, $q(k)\geq e^{-4t}$,
so $E(q+u)$ is bounded near $u=0$. For $u\geq1$, Parseval's
identity and
$|\partial_k\log[w(k)+u]|\leq|w'(k)|/u$, for $w=n,q$,
give a bound proportional to $u^{-2}$. Together,
\begin{equation}
 E(n+u)+E(q+u)\leq C_t(1+|\log u|)\quad(0<u\leq1),\qquad
 E(n+u)+E(q+u)\leq C_t/u^2\quad(u\geq1).
 \label{eq:creation-integrable-szego-bound}
\end{equation}
Monotone convergence in
Eq.~\eqref{eq:creation-entropy-boundary-integral} proves
Eq.~\eqref{eq:creation-entropy-edge-limit}, with
\begin{equation}
 \gamma(t)=\int_0^\infty du\,[E(n+u)+E(q+u)]<\infty.
 \label{eq:creation-boundary-constant}
\end{equation}

For consecutive intervals,
$I(A:C|B)=S_{a+b}+S_{b+c}-S_b-S_{a+b+c}$.
All four lengths tend to infinity when $b\to\infty$, so
Eq.~\eqref{eq:creation-entropy-edge-limit} makes their remainder
terms vanish. This proves the fixed-proportion limit in the main
text, and more generally gives vanishing CMI as $b\to\infty$
for any choices of the end lengths $a,c\geq1$.

For the slab geometry, take $a,c\to\infty$ first, keeping $b$
fixed. The same entropy formula gives
\begin{equation}
 I_{\rm slab}(b;t)
 :=\lim_{a,c\to\infty}I(A:C|B;t)
 =b\,s(t)+\gamma(t)-S_b(t).
 \label{eq:slab-limit}
\end{equation}
Applying Eq.~\eqref{eq:creation-entropy-edge-limit} once more,
now to $S_b$, shows $I_{\rm slab}(b;t)\to0$ as $b\to\infty$.
The auxiliary entropy is extensive and cancels from every CMI
here, so both limits hold for consecutive complete-cell intervals
in the spin construction as well. 

\section{CMI in a spin chain with simple jumps}
\label{sec:mechanism}
\label{sec:response}

The discussion in Sec.~\ref{sec:discussion} concerns a second spin
model, distinct from the auxiliary construction of Sec.~\ref{app:spins}.
Start from $\ket\Omega=\ket{\uparrow\uparrow\cdots}$ and evolve under
\begin{equation}
 \dot\rho=\sum_j\D{J_j}(\rho),\qquad J_j=X_j+X_{j+1}.
 \label{eq:spin-model}
\end{equation}
A jump creates the same superposition of neighboring excitations
as in the fermionic model, so the two models have the same initial
CMI response. Their finite-time states are different,
and the spin data below do not establish an algebraic tail at
fixed positive time.

\subsection{The short-time reduced state}
\label{app:spin-expansion}

Let $\ket{0_L}=\ket{\uparrow}^{\otimes L}$ denote the initial state
of the interval, $\ket{j}=X_j\ket{0_L}$ a single flip, and
$\ket{j,j+1}=X_jX_{j+1}\ket{0_L}$ two neighboring flips. For an
internal bond, the jump term gives
$(\ket{j}+\ket{j+1})(\bra{j}+\bra{j+1})$. Using
\mbox{$(X_j+X_{j+1})^2=2\id+2X_jX_{j+1}$}, the anticommutator contributes
$-2\ket{0_L}\bra{0_L}$ and
$-\ket{j,j+1}\bra{0_L}-\ket{0_L}\bra{j,j+1}$.

For a jump crossing a boundary, the exterior spin is initially polarized
along $Z$, so its expectation of $X$ is zero. Tracing it out leaves
$\D{X_1}(\ket{0_L}\bra{0_L})$ at the left endpoint and the
corresponding expression with $X_L$ at the right endpoint. Jumps wholly
outside the interval make no contribution to its first-order reduced
state. Thus
\begin{equation}
 \rho_L(t)=(1-2Lt)\ket{0_L}\bra{0_L} +t\sum_{i,j=1}^L(K_L)_{ij}\ket{i}\bra{j} -t\sum_{j=1}^{L-1} \bigl(\ket{j,j+1}\bra{0_L}+\ket{0_L}\bra{j,j+1}\bigr) +O_L(t^2),
 \label{eq:spin-rho-expansion}
\end{equation}
where $(K_L)_{ij}=2\delta_{ij}+\delta_{i,j+1}+\delta_{i,j-1}$. Both boundary-crossing
jumps are included even when $L=1$.

At $t=0$, the eigenvalue of $\ket{0_L}$ is separated by one from
the remaining eigenvalues, which are zero. The matrix elements connecting it to
two-flip states therefore affect the eigenvalues only at second order.
Within the initially zero eigenspace, the only nonzero first-order
matrix is $K_L$ on the one-flip states. Its normalized eigenvectors have
components $v_q(j)=\sqrt{2/(L+1)}\sin[\pi qj/(L+1)]$, giving
$\kappa_q=2+2\cos[\pi q/(L+1)]$. The eigenvalues sum to $2L$.
Expanding the entropy gives
\begin{equation}
 S_L(t)=2Lt(1-\log t)+t h_L+O_L(t^2|\log t|),
 \label{eq:spin-entropy-expansion}
\end{equation}
with the same $h_L=-\sum_q\kappa_q\log\kappa_q$ as in
Sec.~\ref{app:fermion-response}. Eigenvalues that first appear at
order $t^2$ contribute at most $O_L(t^2|\log t|)$. The extensive
term cancels from CMI, giving
\begin{equation}
 I(A:C|B;t)=t\chi_{a,c}(b)+O_{a,b,c}(t^2|\log t|),\qquad \chi_{a,c}(b)=h_{a+b}+h_{b+c}-h_b-h_{a+b+c}.
 \label{eq:spin-linear-response}
\end{equation}
The calculation in Sec.~\ref{app:response-asymptotics} therefore gives
\begin{equation}
 \chi_{a,c}(b)=\frac{3\zeta(3)ac}{b^4}+O_{a,c}(b^{-5}).
\end{equation}
Here $t\to0^+$ is taken first, at fixed intervals. The short-time
remainder need not be uniform in $b$, so this result alone does
not determine the tail at fixed $t>0$.

\begin{figure*}[t]
 \centering
 \includegraphics[width=0.96\textwidth]{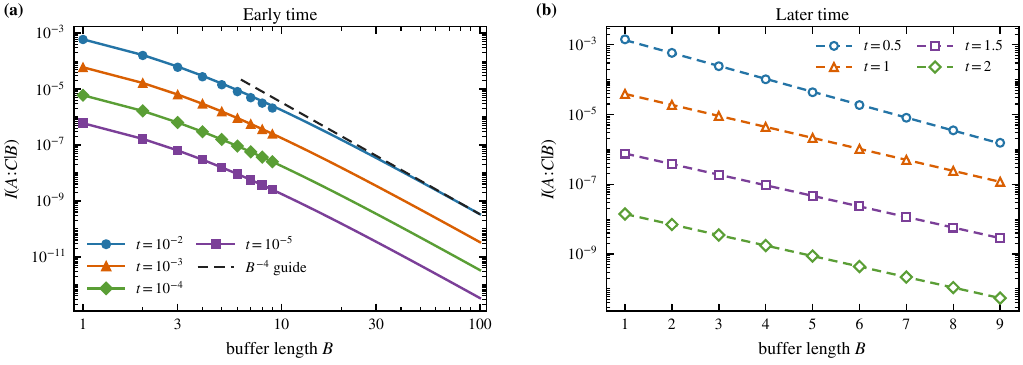}
 \caption{CMI in the $X_j+X_{j+1}$ spin chain, with single-site
end regions. (a) Early-time data approach the linear-response
prediction as $t$ decreases. Curves extending to larger buffers
show the perturbative response and its $b^{-4}$ behavior.
 (b) At later times the accessible buffers show approximately
exponential decay. These finite sizes do not determine the
asymptotic behavior at fixed positive time. The buffer length,
denoted $B$ in the plot, is $b$ in the text.}
 \label{fig:spin-data}
\end{figure*}

\subsection{Finite-time density matrix}
\label{app:spin-numerics}

We obtain the finite-time reduced state without numerically
integrating a master equation. All jumps are diagonal in the $X$ basis.
For interval configurations $s,s'\in\{\pm1\}^L$, write
$\Delta_j=s_j-s'_j$. An internal bond contributes
$\exp[-t(\Delta_j+\Delta_{j+1})^2/2]$ to the decay of the matrix
element. Tracing the exterior equates its bra and ket configurations,
so the two boundary-crossing bonds contribute
$\exp[-t\Delta_1^2/2]$ and $\exp[-t\Delta_L^2/2]$.
The initial polarized state has every $X$-basis matrix element equal
to $2^{-L}$. Hence
\begin{equation}
 \langle s|\rho_L(t)|s'\rangle_X =2^{-L}\exp\!\biggl[-\frac t2 \biggl(\Delta_1^2+\Delta_L^2 +\sum_{j=1}^{L-1}(\Delta_j+\Delta_{j+1})^2\biggr)\biggr].
 \label{eq:spin-exact-density}
\end{equation}
The endpoint terms are both present for $L=1$. This is the reduced
state of an interval in the infinite chain, including the effect of
the adjacent bonds.

Diagonalizing this matrix for the four interval sizes gives the
finite-time CMI compared with the initial response in
Fig.~\ref{fig:spin-data}. The long-buffer response curves can be
computed directly from Eq.~\eqref{eq:spin-h-exact}.

Figure~\ref{fig:spin-data}(a) shows the finite-time CMI approaching
the initial response as $t$ decreases. Direct density-matrix
calculations reach $b=9$, while Eq.~\eqref{eq:spin-h-exact}
allows much larger buffers for the response. At later times,
panel (b) shows approximately exponential decay over the accessible
sizes. Whether this persists asymptotically or is a finite-size
crossover remains open.

\section{Surface-code decoding with correlated spin noise}
\label{app:surface-recovery}

The surface-code calculation discussed in Sec.~\ref{sec:discussion}
tests how a particular decoder responds
to noise with the same $2+2\cos k$ covariance as the spin
model in Sec.~\ref{sec:response}. The input is an encoded state on a two-dimensional
patch, rather than a product state on a chain. The calculation
concerns decoding and does not establish the CMI of this state.

\subsection{Noise on an open patch}

Consider a rotated planar surface code of odd distance $d$, with
data qubits $(r,c)\in\{0,\ldots,d-1\}^2$. We use the standard
surface-code encoding~\cite{DennisKitaevLandahlPreskill2002} and
place jumps on open horizontal bonds:
\begin{equation}
 \cL_d=\sum_{r=0}^{d-1}\sum_{c=0}^{d-2}\D{J_{r,c}},
 \qquad J_{r,c}=Z_{r,c}+Z_{r,c+1}.
 \label{eq:surface-jumps}
\end{equation}
No bond joins the two ends of a row or extends outside the patch.
For a Hermitian $J$, the dissipator is
$\D{J}=-[J,[J,\cdot]]/2$, so its evolution is an average of
$e^{-i\phi J}$ with $\phi$ Gaussian of variance $t$. All the
bond jumps commute. Drawing independent
$\phi_{r,c}\sim\mathcal N(0,t)$ therefore gives the exact noise
channel
\begin{equation}
 \rho_d(t)=\mathbb E_\phi[U(\phi)\rho_{\rm SC}U(\phi)^\dagger],\qquad U(\phi)=\exp\!\left[-i\sum_{r,c}\vartheta_{r,c}Z_{r,c}\right],\qquad \vartheta_{r,c}=\phi_{r,c-1}+\phi_{r,c},
 \label{eq:surface-angles}
\end{equation}
where $\phi_{r,-1}=\phi_{r,d-1}=0$.
The angle variance is $t$ at the end columns and $2t$ in the
bulk; neighboring angles in one row have covariance $t$, and
all other covariances vanish. Thus the bulk covariance has
symbol $t(2+2\cos k_x)$ despite being strictly finite range.
On each open row, the alternating combination
$\sum_c(-1)^c\vartheta_{r,c}=0$ vanishes for every realization.

\subsection{Logical error after full-patch decoding}

For each sample, we draw a bond field from
Eq.~\eqref{eq:surface-angles}. A site-dependent extension of the
fermionic simulation method of
Ref.~\cite{BravyiEnglbrechtKoenigPeard2018} then samples the
syndrome and evaluates the residual logical rotation after
equal-weight MWPM~\cite{Higgott2022PyMatching}. The decoder is
given the syndrome, not the sampled field. Equal weights define
a fixed benchmark for this anisotropic, correlated noise.

Let $\vartheta_L(\phi,s)$ be the remaining logical rotation for
field $\phi$ and syndrome $s$. After averaging over both and discarding
the syndrome record, the logical channel is an average of
$e^{i\vartheta_L Z_L}$. Its off-diagonal matrix element is
multiplied by $g_L=\mathbb E_{\phi,s}e^{2i\vartheta_L}$, giving
\begin{equation}
 D^L_{\diamond,{\rm av}}(d,t)
 =\|\mathcal E_L^{(d,t)}-\mathcal I_L\|_\diamond
 =|1-g_L|.
 \label{eq:surface-logical-error}
\end{equation}
We average before taking the channel norm because the field is an
unobserved noise variable.

\begin{figure*}[t]
 \centering
 \includegraphics[width=0.94\textwidth]
 {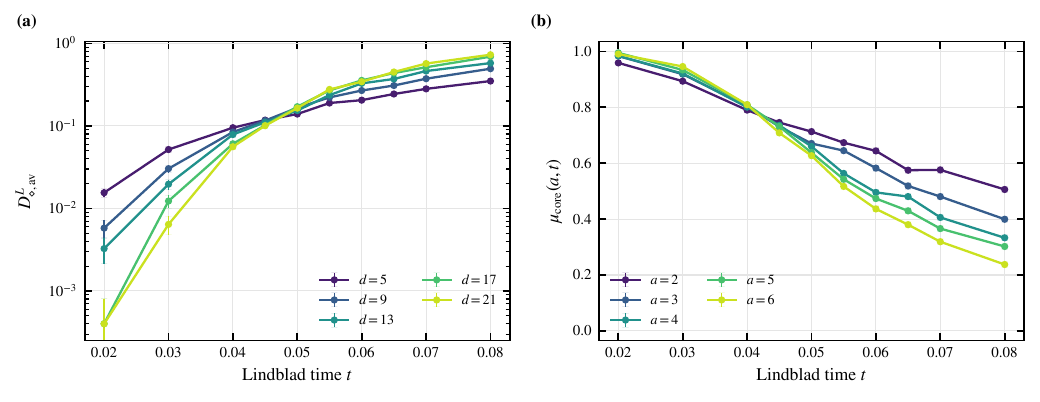}
 \caption{Surface-code decoding under the correlated noise of
 Eq.~\eqref{eq:surface-jumps}. (a) Logical-channel distance from
 the identity, without a factor of one half, after full-patch
 minimum-weight perfect matching (MWPM), for code distances
 $d=5,9,13,17,21$. (b) Probability that buffered and full-patch
 MWPM assign the same partner to every syndrome defect in the
 central core, for core half-width and buffer width
 $a=2,3,4,5,6$. The corresponding patch distances are
 $d=11,15,21,25,31$. Error bars show one standard error; lines
 guide the eye. At the shortest sampled times both diagnostics
 improve overall as their respective geometries grow. }
 \label{fig:surface-code-gapless-decoding}
\end{figure*}

\subsection{Decoding inside a buffered window}

The second test asks how much syndrome information is needed to
reproduce the central matching decisions of the global decoder.
A central core has nominal half-width $a$ and is surrounded by
a buffer of width $a$, so the matching window has half-width
$2a$. We compare MWPM inside that window with MWPM on the
whole patch. For each syndrome defect $x$ in the core, denote
the two assigned partners by $m_{\rm loc}(x)$ and
$m_{\rm glob}(x)$. Writing $B_s$ for the set of core defects,
the plotted probability is
\begin{equation}
 \mu_{\rm core}(a,t)=\Pr_{\phi,s}\!\left[
 m_{\rm loc}(x)=m_{\rm glob}(x)
 \quad\forall x\in B_s\right].
 \label{eq:surface-core-agreement}
\end{equation}
This comparison includes partners in the buffer. It distinguishes
matching to the artificial window boundary from matching to a
physical boundary of the code.

\end{document}